\documentclass[12pt,letterpaper]{article}

\usepackage{amsmath}
\usepackage{amsfonts}
\usepackage{amssymb}
\usepackage{makeidx}
\usepackage{graphicx}
\usepackage[left=1.00in, right=1.00in, top=1.00in, bottom=1.00in]{geometry}
\usepackage{color}
\usepackage{setspace}
\usepackage{float}
\restylefloat{table}
\usepackage{hyperref, longtable, pdflscape}
\usepackage{appendix}
\usepackage{titlesec}
\titleformat*{\section}{\large\bfseries}
\titleformat*{\subsection}{\normalsize\bfseries}

\usepackage{psfrag,epsf}
\usepackage{enumerate}
\usepackage{natbib}
\usepackage{url} 

\usepackage{tikz, amssymb, algorithm, algpseudocode}

\usepackage{refcount}
\usepackage{amsthm}
\newtheorem{assumption}{Assumption}
\newtheorem{theorem}{Theorem}

\newtheorem*{assumptstar}{Assumption \assmpstarnumber} 
\makeatletter
\newenvironment{assumption*}[1]
{\edef\assmpstarnumber{\getrefnumber{#1}*}%
	\let\@currentlabel\assmpstarnumber\assumptstar}
{\endassumptstar}
\makeatother

\theoremstyle{remark}
\newtheorem{remark}{Remark}

\newcommand{\indept}{\mathrel{\text{\scalebox{1.07}{$\perp\mkern-10mu\perp$}}}} 

\usepackage{array}
\newcolumntype{P}[1]{>{\centering\arraybackslash}p{#1}} 

\begin{document}
\begin{center}
	{\Large Inverse Conditional Probability Weighting with Clustered Data in Causal Inference}\\ 
	\vspace{0.2in}
	{ Zhulin He}\\
	\today
\end{center}

\bigskip
\begin{abstract}
	Estimating the average treatment causal effect in clustered data often involves dealing with unmeasured cluster-specific confounding variables. Such variables may be correlated with the measured unit covariates and outcome. When the correlations are ignored, the causal effect estimation can be biased. By utilizing sufficient statistics, we propose an inverse conditional probability weighting (ICPW) method, which is robust to both (i) the correlation between the unmeasured cluster-specific confounding variable and the covariates and (ii) the correlation between the unmeasured cluster-specific confounding variable and the outcome. Assumptions and conditions for the ICPW method are presented. We establish the asymptotic properties of the proposed estimators. Simulation studies and a case study are presented for illustration.
\end{abstract}
	
\noindent%
 \vfill	
	
\newpage

\section{Introduction}
\label{s:intro}

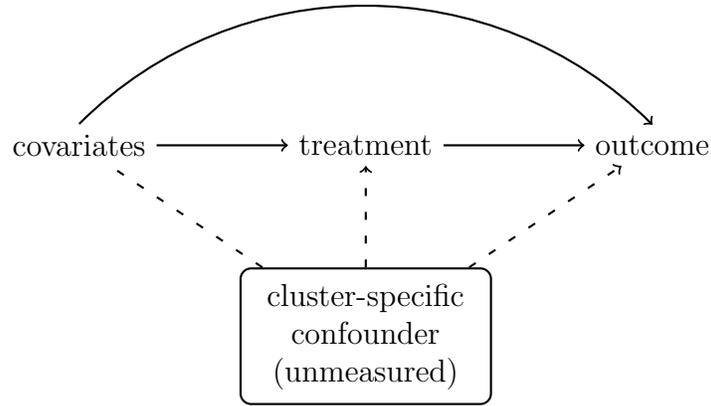
\begin{figure}
	\centering
	\makebox{
		\begin{tikzpicture}[node distance = 1.5in]
		\node(X){covariates};
		\node[right of = X, yshift = -.0cm](A){treatment};
		\node[right of = A, yshift = -.0cm](Y){outcome};
		\node[draw=black, thick, rounded corners, below of = A, yshift = 0.5in](U){\begin{tabular}{c} cluster-specific\\confounder\\ (unmeasured)\end{tabular}};
		\draw [->, thick] (X) -- (A);
		\draw [->, thick] (A) -- (Y);
		\draw [->, thick] (X.north) to [out=45,in=135] (Y.north);	
		\draw [-, thick, loosely dashed] (U) -- (X);
		\draw [->, thick, loosely dashed] (U) -- (A);
		\draw [->, thick, loosely dashed] (U) -- (Y);
		\end{tikzpicture}
	}
	\caption{\label{fig:intro1}A graph illustrating possible relationships between unmeasured cluster-specific confounder and measured variables.}
\end{figure}

Clustered data are usually considered as groups of units that share the same or similar characters. Some examples of clustered data are children in the classes or schools, family members in the households, and animals in the feedlots or barns. Estimating the average treatment causal effect in clustered data often involves dealing with unmeasured pre-treatment cluster-specific confounding variables, which can bring challenges in the estimation procedures. The cluster-specific confounding variables in the previous examples can be teachers' experience and school resource \citep[e.g.,][]{hong2006evaluating}, neighborhood environment for the households \citep[e.g.,][]{brumback2011adjusting}, and management and operations in the feedlots or barns \citep[e.g.,][]{o2005association, ramirez2012efficient}. There are two possible reasons for why such variables are not collected into data. The first possible reason, from data collection point of view, is that it may be difficult or impossible to measure a cluster-specific confounding variable. The second possible reason, from estimation point of view, is that the cluster-specific confounding variable may be not of interest in estimation. Usually when the cluster-specific confounding variable is unobserved, its relationship to other measured variables may be unclear, which can result biased causal effect estimates. As shown in Figure \ref{fig:intro1}, a dashed  line or a dashed arrow represents an unclear relationship between two variables. When all three kinds of relationships with respect to unmeasured cluster-specific confounding variable are unknown, it is impossible for us to adjust for this unmeasured cluster-specific confounding variable. Therefore, additional assumptions are needed for the adjustment.

One assumption we can consider is that the treatment assignment mechanism is known. This implies the relationship between the unmeasured cluster-specific confounder and the treatment is known. Then the corresponding arrow from ``cluster-specific confounder" to ``treatment" in Figure \ref{fig:intro1} is not dashed anymore. Under such assumption, the inverse probability weighting (IPW) or inverse propensity score weighting, an important tool used in causal inference, can be applied to both randomized experiments and observational studies. A general introduction of IPW method in causal inference can be found in \citet[Section 2.4]{hernan2018causal}. The IPW method involves estimating the probability, which is also known as a propensity score \citep{rosenbaum1983central}, of a unit being applied the treatment given some information. The method has been widely studied in causal inference \citep[e.g.,][]{robins2000marginal, hirano2001estimation, lunceford2004stratification, cole2008constructing, vanderweele2009marginal, ertefaie2010comparing, tan2010bounded, sjolander2011comparing, zhang2012robust, tchetgen2012causal, vansteelandt2014regression, imai2014covariate, naimi2014constructing, austin2015moving, ogburn2015doubly, liu2016inverse}, missing data analysis  (e.g., \citealp{little1986survey, rotnitzky1995semiparametric, hogan2004handling}; \citealp[Chapter 6]{tsiatis2006semiparametric}; \citealp{chen2008improving, kott2010using}; \citealp[Chapter 5]{kim2013statistical}; \citealp{mitra2011estimating, miao2015identification, sun2017inverse, ding2018causal, wen2018semi}), and survey statistics \citep[e.g.,][]{deville1992calibration, fuller1994regression, kalton2003weighting, kim2014propensity}. An early case of IPW dates back to the Horvitz-Thompson estimator \citep{horvitz1952generalization}, where the probability of a binary indicator for sampling (or missingness) is used for estimation. 

The IPW method usually requires all information for estimation, which is challenging for clustered data when cluster-level information is partially or completely missing. In such case, the assumption of no unmeasured confounder is violated. Without considering the existence of the unmeasured confounder, the method can lead researchers to the Simpson's paradox \citep{simpson1951interpretation}. This is described by \citet[Section 3.6]{pearl2016causal}. Sensitivity analyses of the IPW estimators, when no unmeasured confounder assumption is violated, has been studied \citep[see][]{brumback2004sensitivity, zhao2017sensitivity}. 

Efforts, using the IPW method, to adjust for the unmeasured cluster-specific confounding variable have been made in recent years. \cite{li2013propensity} treated the unmeasured cluster-specific confounding variable as random effect and fixed effect in two propensity score models, among several models they considered, to investigate the performance of the propensity score weighting methods. As discussed by \cite{li2013propensity}, when the number of clusters is large and the cluster size is small, fixed effect model can lead to unstable propensity score estimates due to the Neyman-Scott incidental parameter problem \citep{neyman1948consistent}. Comparatively, the random effect model does not have such problem, but it requires the independence between covariates and the unmeasured cluster-specific confounding variable. However, the independence requirement can not always be guaranteed. \cite{yuan2007model} showed biased estimation, in a missing data setting, when the outcome depends on the unmeasured cluster-specific confounding variable which may be correlated with the covariates. \cite{skinner2011inverse} proposed an IPW method using conditional logistic regression to overcome the bias caused by aforementioned correlations. Their method was originated from a missing data setting, then extended to binary treatment effect estimation. Later, \cite{yang2017propensity} developed calibrated propensity scores for binary treatment effect estimation, which is robust to model misspecification. Other methods using propensity score with clustered data are matching \citep[e.g.,][]{arpino2011specification, arpino2016propensity, zubizarreta2017optimal} and stratification \citep[e.g.,][]{thoemmes2011use}. 

In this paper, we focus on a novel method when the cluster-level confounding variable is unobserved. By utilizing the sufficient statistics, we proposed an inverse conditional probability weighting (ICPW) method, which is robust to both (i) the correlation between the unmeasured cluster-specific confounding variable and the covariates (i.e., the left dashed line in Figure \ref{fig:intro1}) and (ii) the correlation between the unmeasured cluster-specific confounding variable and the outcome (i.e., the right dashed arrow in Figure \ref{fig:intro1}). 

The remainder of this paper is arranged as follows. Section \ref{sec:setup} describes clustered data structure, assumptions and models. In Section \ref{sec:prop.method}, we propose the ICPW method by utilizing the sufficient statistics. Asymptotic properties of the proposed estimators are shown in Section \ref{sec:asymp}. Simulation studies and a case study are conducted in Section \ref{sec:simu} and Section \ref{sec:case}, respectively. We conclude the paper with discussion in Section \ref{sec:discuss}.

\section{Basic Setup}\label{sec:setup}
\subsection{Clustered Data Structure and Estimand of Interest}\label{subsec2:cluster}
Let $ Y_{ij} $ be the observed outcome for the $ j $th unit ($ j=1,2,\ldots,n_i $) in the $ i$th cluster ($ i=1,2,\ldots,m $). Denote by a $p$-dimensional vector $ \boldsymbol{X}_{ij} $ the observed unit-specific pre-treatment covariates. Let $ A_{ij} $ be the treatment variable with domain $ \Omega_{A} $. 
For categorical treatments, we index treatment levels by a series of integers 0 to $ K $, where $ K\geq 1 $. Assume there is no hidden variations of treatments, which is one component of the stable unit treatment value assumption (SUTVA) \citep[Section 1.6]{imbens2015causal}. Denote the sample size by $ n=\sum_{i=1}^{m}n_i $. For cluster-level notations, let $ \boldsymbol{Y}_i=(Y_{i1},\ldots,Y_{in_i})^T $, $ \boldsymbol{X}_i=(\boldsymbol{X}_{i1}^T,\ldots,\boldsymbol{X}_{in_i}^T)^T $, and $ \boldsymbol{A}_i=(A_{i1},\ldots,A_{in_i})^T $ be the $ i $th cluster-level outcome, covariate, and treatment indicator, respectively. Also, let $  \boldsymbol{U}_i $ be a cluster-specific confounding variable summarizing unobserved information of cluster-level confounders. Assume $ \Omega_{U} $, the domain of $  \boldsymbol{U}_i $, is compact.

Next, we follow the potential outcome (or called counterfactual) setup \citep{rubin1974estimating, splawa1990application}. Suppose each unit has two potential outcomes, $ Y_{ij}(0) $ and $ Y_{ij}(1) $. In particular, $ Y_{ij}(0) $ is the outcome that would be realized, if the unit received control, and $ Y_{ij}(1) $ is the outcome that would be realized, if the unit received treatment. Denote cluster-level potential outcomes as $ \boldsymbol{Y}_{i}(0)=(Y_{i1}(0),\ldots,Y_{in_i}(0))^T $ and $ \boldsymbol{Y}_{i}(1)=(Y_{i1}(1),\ldots,Y_{in_i}(1))^T $. More generally, denote cluster-level potential outcome with treatment level $ a $ as $ \boldsymbol{Y}_{i}(a)=(Y_{i1}(a),\ldots,Y_{in_i}(a))^T $, where $ Y_{ij}(a) $ is the unit potential outcome.

Our goal for binary treatment is to estimate the population average treatment effect, $ \tau=E\{Y(1)-Y(0)\} $, which is the expectation of difference between two potential outcomes over the population. There are two ways to estimate $ \tau $ without modeling potential outcomes. The first one is to calculate the unit treatment effect, namely, $ Y(1)-Y(0) $, and then take the expectation with respect to the population. However, this method is not feasible due to the fundamental problem of causal inference \citep{rubin1974estimating, holland1986statistics}. Specifically, each unit can receive either treatment or control, so only one of the potential outcomes can be observed. Therefore, the unit treatment causal effect can not be directly calculated, which implies the first way does not work. The second way for $ \tau $ estimation is first taking the expectations of both potential outcomes over the population, namely, $ E\{Y(0)\} $ and $ E\{Y(1)\} $, and then calculating the difference of the two expectations for $ \tau $. Such estimand is proposed in \cite{rosenbaum1983central}. We also consider the latter one in the  paper. 

For a general notation, we are interested in estimating $ E\{Y(a)\} $ and $ E\{Y(a')\} $ with treatment levels $ a $ and $ a' $, where $ a\neq a' $. Then the causal effect can be constructed as a function of $ E\{Y(a)\} $ and $ E\{Y(a')\} $. For example, the causal risk difference, causal relative risk, and causal odds ratio for binary outcome can be constructed as $ P\{Y(a)=1\}-P\{Y(a')=1\} $, $ P\{Y(a)=1\}/P\{Y(a')=1\} $, and 
$
\frac{P\{Y(a)=1\}}{1- P\{Y(a)=1\}}\Big/\frac{P\{Y(a')=1\}}{1-P\{Y(a')=1\}},
$
respectively, for $ a\neq a' $. In such case, we are interested in estimating $ P\{Y(a)=1\} $  and $ P\{Y(a')=1\} $.

\subsection{Assumptions and Propensity Score for Inverse Probability Weighting}\label{sebsec2:assumptions}
In order to identify the population average treatment effect, we consider some assumptions hold in the clustered data. Usually most assumptions in causal inference are listed in unit level. However, clustered data is different in data structure. To emphasize such difference, we consider the following assumptions (except Assumption \ref{assum:consistency1}) in cluster level. Besides, all assumptions (except Assumption \ref{assum:positivity1}) are listed with respect to binary treatment. The corresponding general forms for non-binary treatments are given in the immediate discussion. 

\begin{assumption}\label{assum:cluster}
	$ \{\boldsymbol{A}_i, \boldsymbol{X}_i, \boldsymbol{Y}_i(0), \boldsymbol{Y}_i(1), \boldsymbol{U}_i\} \indept \{\boldsymbol{A}_{i'}, \boldsymbol{X}_{i'}, \boldsymbol{Y}_{i'}(0), \boldsymbol{Y}_{i'}(1), \boldsymbol{U}_{i'} \} $ for any $ i\neq i' $. Moreover, $ A_{ij} \indept A_{ij'}|\boldsymbol{X}_i, \boldsymbol{U}_i $ for all clusters and $ j\neq j' $.
\end{assumption}
The first component in Assumption \ref{assum:cluster} assumes all clusters are independent of each other. It satisfies the ``no interference" component in the SUTVA assumption \citep[Section 1.6]{imbens2015causal} in cluster level. That means the treatments applied to the units in one cluster do not affect the potential outcomes of the units in any other clusters. A more general form of the first component is $ (\boldsymbol{A}_i, \boldsymbol{X}_i, \{\boldsymbol{Y}_i(a)\}_{a\in\Omega_A},  \boldsymbol{U}_i) \indept (\boldsymbol{A}_{i'}, \boldsymbol{X}_{i'}, \{\boldsymbol{Y}_{i'}(a)\}_{a\in\Omega_A},  \boldsymbol{U}_{i'} ) $.

The second component in Assumption \ref{assum:cluster} describes the conditional independence of the treatment assignment mechanism for units within one cluster. That is, given all information of covariates $ \boldsymbol{X}_i $ and confounding variable $  \boldsymbol{U}_i $ in the cluster, treatment applied to one unit does not affect that applied to other units within the same cluster.

\begin{assumption}[Consistency]\label{assum:consistency1}
	$ Y_{ij} = Y_{ij}(0) \mathcal{I}\{A_{ij}=0\} + Y_{ij}(1) \mathcal{I}\{A_{ij}=1\} $, for all $ i $ and $ j $.
\end{assumption}
Assumption \ref{assum:consistency1} sets up the linkage between observed outcome and potential outcomes for each unit \citep[Section 1.1]{hernan2018causal}. The meaning of this assumption is straightforward. If one unit receives control, then potential outcome $ Y_{ij}(0) $ is observed. Similarly, if one unit receives treatment, then potential outcome $ Y_{ij}(1) $ is observed. A more general description of the consistency assumption is that if $ A_{ij}=a\in\Omega_{A} $, then $ Y_{ij} = Y_{ij}(a) $.

\begin{assumption}[Cluster-level Positivity]\label{assum:positivity1} The cluster-level treatment joint probability is $ P(\boldsymbol{A}_i=\boldsymbol{a}_i|\boldsymbol{X}_i, \boldsymbol{U}_i)=\prod_{j=1}^{n_i}P(A_{ij}=a_{ij}|\boldsymbol{X}_{i}, \boldsymbol{U}_i) $. It satisfies
	$ 0< P({A}_{ij}={a}_{ij}|\boldsymbol{X}_i, \boldsymbol{U}_i)<1 $, for all $ i $, $ j $, and $ \boldsymbol{a}_i=(a_{i1},\ldots,a_{in_i}) $, with $ a_{ij}\in\Omega_{A} $. When the treatment is binary, all elements in $ \boldsymbol{a}_i $ are binary, and $ \boldsymbol{a}_i\neq \boldsymbol{0} $ or $ \boldsymbol{1} $.
\end{assumption}
The unit-level Positivity assumption for binary treatment is $ 0<P(A_{ij}=0|\boldsymbol{X}_{i}, \boldsymbol{U}_i)<1 $ and $ 0<P(A_{ij}=1|\boldsymbol{X}_{i}, \boldsymbol{U}_i)<1 $. It is not equivalent to Assumption \ref{assum:positivity1} because of the constraint $ \boldsymbol{a}_i\neq \boldsymbol{0} $ or $ \boldsymbol{1} $. Such constraint excludes those clusters that all units in one cluster only received treatment (or control). Besides, the equivalence $P(\boldsymbol{A}_i=\boldsymbol{a}_i|\boldsymbol{X}_i, \boldsymbol{U}_i)=\prod_{j=1}^{n_i}P(A_{ij}=a_{ij}|\boldsymbol{X}_{i}, \boldsymbol{U}_i) $ is obtained from the second component in Assumption \ref{assum:cluster}.

\begin{assumption}[Cluster-level Ignorability]\label{assum:ignorability1}
	$ \{\boldsymbol{Y}_{i}(0), \boldsymbol{Y}_{i}(1)\} \indept \boldsymbol{A}_{i} | \boldsymbol{X}_i,  \boldsymbol{U}_i $ for all $ i $.
\end{assumption}
Assumption \ref{assum:ignorability1} indicates that, in each cluster, all units' treatment assignments are not affected by the units' potential outcomes given information of $ \boldsymbol{X}_i $ and $  \boldsymbol{U}_i $. It is different from the another form of Ignorability assumption, $ \{\boldsymbol{Y}_{i}(0), \boldsymbol{Y}_{i}(1)\} \indept \boldsymbol{A}_{i} | \boldsymbol{X}_i $, which indicates no unmeasured confounder. For clustered data, cluster-level confounding factors may be various across clusters. Their existence should not be ignored. Instead, Assumption \ref{assum:ignorability1} allows the existence of unmeasured cluster-level confounding variable. A more general form of Assumption \ref{assum:ignorability1} is $ \{\boldsymbol{Y}_{i}(a)\}_{a\in\Omega_{A}} \indept \boldsymbol{A}_{i} | \boldsymbol{X}_i,  \boldsymbol{U}_i $ for all $ i $.

Under the aforementioned assumptions, for binary treatment, the IPW estimator for the average treatment effect is expressed as
\begin{equation}\label{eq:ipw_tau} 
{\tau}_{IPW}=\frac{1}{n}\sum_{i}^{m}\sum_{j}^{n_i}\Big\{\frac{A_{ij}Y_{ij}}{P(A_{ij}=1|\boldsymbol{X}_{i},U_i)} - \frac{(1-A_{ij})Y_{ij}}{1-P(A_{ij}=1|\boldsymbol{X}_{i},U_i)} \Big\},
\end{equation}
where the propensity score $ P(A_{ij}=1|\boldsymbol{X}_{i},U_i) $ is the conditional probability of being applied the treatment given $ (\boldsymbol{X}_{i}, U_i) $. In applications, model for unit-level treatment indicator $ A_{ij} $ can be constructed using a generalized linear mixed effect model
\begin{equation}\label{eq:glm_A_ij}
P(A_{ij}=a|\boldsymbol{X}_{i},U_i)=g(\boldsymbol{X}_{ij}^T\boldsymbol{\beta}+U_i),
\end{equation}
for all $ a $, where  $ g $ is the link function, and $ \boldsymbol{\beta} $ is a $ p $-dimensional vector of parameter. For binary treatment indicator, researchers usually choose logic link as the link function. Then we have the following form of a logistic model,  
\begin{equation}\label{eq:logistic_A_ij}
P(A_{ij}=1|\boldsymbol{X}_{ij},U_i)=\frac{\exp(\boldsymbol{X}_{ij}^T\boldsymbol{\beta}+U_i)}{1+\exp(\boldsymbol{X}_{ij}^T\boldsymbol{\beta}+U_i)}.
\end{equation}

For multiple treatments, denote by $k$ the treatment level with range $k=0,\dots,K$, where $K\leq 1$. Therefore, there are $K+1$ treatment levels in total. Assume treatment assignment follows a multinomial logistic model. That is,
\begin{align}\label{eq:multiple_trt}
P(A_{ij}=a|\boldsymbol{X}_{ij},\boldsymbol{U}_{i})= \dfrac{\exp\{\sum_{k=1}^{K}\mathcal{I}(a=k)(\boldsymbol{X}_{ij}^T\boldsymbol{\beta}_k+U_{ik})\}+\mathcal{I}(a=0)}{1+\sum_{h=1}^{K}\exp(\boldsymbol{X}_{ij}^T\boldsymbol{\beta}_h+U_{ih})},
\end{align}
where $ a=0,\dots,K $, $\boldsymbol{\beta}_k$ is the parameter for $k$th treatment assignment and $U_{ik}$ is the unmeasured cluster-specific variable for the $k$th treatment. Then we have the cluster-specific confounding variable as $\boldsymbol{U}_i=(U_{i1},\dots,U_{iK})$ for the $i$th cluster, with dimension $K$. 

We should note that propensity score formulas above involve with the knowledge of $ \{\boldsymbol{U}_i\}_{i=1}^{m} $, which is unobserved in data. Besides, the existence of  unmeasured $ \{\boldsymbol{U}_i\}_{i=1}^{m} $ is nonignorable. When $ \{\boldsymbol{U}_i\}_{i=1}^{m} $ are treated as fixed effects and estimated by maximizing the overall likelihood, the estimates tends to be biased as the number of cluster $ m $ increases \citep{neyman1948consistent}. Moreover, when $ \{\boldsymbol{U}_i\}_{i=1}^{m} $ are treated as random effects, the requirement of independence between $ \boldsymbol{U}_i $ and $ \boldsymbol{X}_i $ cannot always be guaranteed. So we are motivated to seek an estimation procedure without directly dealing with $ \{\boldsymbol{U}_i\}_{i=1}^{m}  $. Besides, we want to specify under what conditions, the method is feasible.

\subsection{Two Theorems Utilizing Sufficient Statistics}\label{subsec3:theorem}
Before introducing the proposed method, we introduce two theorems utilizing sufficient statistics. These two theorems provide theoretical foundations to our proposed method. In particular, the new method is constructed by utilizing a sufficient statistic in each cluster.

\begin{theorem}\label{thm:prob}
	Suppose $ \boldsymbol{Z} $ and $ \boldsymbol{X} $ are random variables with domain $ \Omega_{\boldsymbol{Z}} $ and $ \Omega_{\boldsymbol{X}} $, and $ \boldsymbol{\theta} $ is a parameter vector with domain $ \Omega_{\boldsymbol{\theta}} $. Let $ \boldsymbol{Z}_{sub}=(Z_{J_1},\ldots,Z_{J_k}) $ with domain $ \Omega_{\boldsymbol{Z}_{sub}} $ be a subvector of $ \boldsymbol{Z}=(Z_{1},\ldots,Z_{n}) $, where $ \{J_1,\ldots,J_k\}\subseteq\{1,\ldots,n\} $. Let $ \boldsymbol{T} $ be a function of $ \boldsymbol{Z} $ with domain $ \Omega_{\boldsymbol{T}} $ satisfies that for each element $ \boldsymbol{t}\in\Omega_{\boldsymbol{T}} $, there exist at least two elements, $ \boldsymbol{z}^o, \boldsymbol{z}^* \in \Omega_{\boldsymbol{Z}} $ and their corresponding subvectors $ \boldsymbol{z}_{sub}^o, \boldsymbol{z}_{sub}^*\in\Omega_{\boldsymbol{Z}_{sub}} $ such that (i) $ \boldsymbol{z}_{sub}^o \neq \boldsymbol{z}_{sub}^* $ and (ii) $ T(\boldsymbol{z}^o)=T(\boldsymbol{z}^*)=\boldsymbol{t} $. If $ \boldsymbol{T} $ is sufficient for $ \boldsymbol{\theta} $, and $0< P(\boldsymbol{Z}=\boldsymbol{z}|\boldsymbol{X},\boldsymbol{\theta})\leq P(\boldsymbol{Z}_{sub}=\boldsymbol{z}_{sub}|\boldsymbol{X},\boldsymbol{\theta})<1 $ for any $ \boldsymbol{z}\in \Omega_{\boldsymbol{Z}} $ and its corresponding subvector $ \boldsymbol{z}_{sub}\in\Omega_{\boldsymbol{Z}_{sub}} $ , then 
	$
	0<P\{\boldsymbol{Z}_{sub}=\boldsymbol{z}_{sub}|\boldsymbol{X},\boldsymbol{T}=T(\boldsymbol{z})\}<1.
	$
\end{theorem}
The proof of Theorem \ref{thm:prob} is in the Supplementary Materials.
\begin{remark}
	Theorem \ref{thm:prob} indicates that by utilizing a sufficient statistic $ \boldsymbol{T} $ for $ \boldsymbol{\theta} $, one can still obtain a non-zero conditional probability of $ \boldsymbol{Z}_{sub} $, which does not depend on $ \boldsymbol{\theta} $ anymore. It is helpful, when one wants to avoid the involvement of nuisance parameter $ \boldsymbol{\theta} $ and maintains the same probability range. Moreover, one should notice that the two probabilities, $ P(\boldsymbol{Z}_{sub}=\boldsymbol{z}_{sub}|\boldsymbol{X},\boldsymbol{\theta}) $ and $ P\{\boldsymbol{Z}_{sub}=\boldsymbol{z}_{sub}|\boldsymbol{X},\boldsymbol{T}=T(\boldsymbol{z})\} $, are not necessarily the same. Besides, the dimensions of $ \boldsymbol{T} $ and $ \boldsymbol{\theta} $ are the same \citep[Section 2.5]{cox2006principles}. One special case of the theorem is setting $ \boldsymbol{Z}_{sub}=\boldsymbol{Z} $. That means we are considering the range of the conditional probability of $ \boldsymbol{Z} $, which is $0<P\{\boldsymbol{Z}=\boldsymbol{z}|\boldsymbol{X},\boldsymbol{T}=T(\boldsymbol{z})\}<1 $.
	
	When applying Theorem $ \ref{thm:prob} $, we have to pay attention to the requirement for the sufficient statistic $ \boldsymbol{T} $, which is stronger than surjection. If $ \boldsymbol{T} $ is a surjective function, it means for any $ \boldsymbol{t}\in\Omega_{\boldsymbol{T}} $ there exists at least one element $ \boldsymbol{z}\in\Omega_{\boldsymbol{Z}} $ and a corresponding subvector $ \boldsymbol{z}_{sub}\in\Omega_{\boldsymbol{T}_{sub}} $ such that $ T(\boldsymbol{z})=\boldsymbol{t} $. In this case, 
	the conclusion in Theorem \ref{thm:prob} is changed to $ 0<P\{\boldsymbol{Z}_{sub}=\boldsymbol{z}_{sub}|\boldsymbol{X},\boldsymbol{T}=T(\boldsymbol{z})\}\leq 1 $. This means if $ \boldsymbol{T} $ is a surjective function, the  probability of $ \boldsymbol{Z}_{sub} $ conditional on $ \boldsymbol{T} $ can be 1, even though the original probability of $ \boldsymbol{Z}_{sub} $ conditional on $ (\boldsymbol{X}, \boldsymbol{\theta}) $ is in range (0,1). In order to make the conditional probability $ P\{\boldsymbol{Z}_{sub}=\boldsymbol{z}_{sub}|\boldsymbol{X},\boldsymbol{T}=T(\boldsymbol{z})\} $ not equal to 1, we have to construct $ \boldsymbol{T} $ more restrictive than surjective. That is, we require at least ``two" elements rather than ``one" element $ \boldsymbol{z}^o,\boldsymbol{z}^*\in\Omega_{\boldsymbol{Z}} $ and corresponding subvectors $ \boldsymbol{z}_{sub}^o,\boldsymbol{z}_{sub}^*\in\Omega_{\boldsymbol{Z}_{sub}} $ such that $ \boldsymbol{z}_{sub}^o\neq \boldsymbol{z}_{sub}^* $ and $ T(\boldsymbol{z}^o)=T(\boldsymbol{z}^*)=\boldsymbol{t} $ for any $ \boldsymbol{t}\in \Omega_{\boldsymbol{T}} $.
\end{remark}

\begin{theorem}\label{thm:indep}
	Suppose $ \boldsymbol{Z}_1 $, $ \boldsymbol{Z}_2 $ and $ \boldsymbol{Z}_3 $ are random variables, and the correspoding domains are $ \Omega_{\boldsymbol{Z}_1} $, $ \Omega_{\boldsymbol{Z}_2} $ and $ \Omega_{\boldsymbol{Z}_3} $, respectively. Let $ \boldsymbol{\theta} $ be a parameter with domain $ \Omega_{\boldsymbol{\theta}} $. Let $ \boldsymbol{T} $, a function of $ \boldsymbol{Z}_1 $, be sufficient for $ \boldsymbol{\theta} $. If $ \boldsymbol{Z}_1\indept \boldsymbol{Z}_2|\boldsymbol{Z}_3,\boldsymbol{\theta} $, then 
	$
	\boldsymbol{Z}_1\indept \boldsymbol{Z}_2|\boldsymbol{Z}_3,\boldsymbol{T}.
	$
\end{theorem}
The proof of Theorem \ref{thm:indep} is in the Supplementary Materials.
\begin{remark}
	Theorem \ref{thm:indep} has great potential in applications when dealing with nuisance parameters, which are nonignorable and not of main interest in estimation. Specifically, when two random variables are independent conditional on a nuisance parameter, one can check whether there exist a sufficient statistic, which is a function of the random variable $ \boldsymbol{Z}_1 $. If such sufficient statistic exists, then a new independence holds, which is conditional on the sufficient statistic rather than the parameter. The new independence is usually more desirable since it only involves with $ (\boldsymbol{Z}_1, \boldsymbol{Z}_2, \boldsymbol{Z}_3, \boldsymbol{T}) $, which are usually formed from data. To obtain the independence conditional on the sufficient statistic via Theorem \ref{thm:indep}, we do not need information on (i) the further requirement of sufficient statistic described in Theorem \ref{thm:prob}, or (ii) the prior distribution of the parameter $ \boldsymbol{\theta} $, or (iii) the relationship between $ \boldsymbol{Z}_2 $ and $ \boldsymbol{\theta} $, or (iv) the relationship between $ \boldsymbol{Z}_3 $ and $ \boldsymbol{\theta} $. It means this theorem has a great property of sufficient statistics in applications. To apply the theorem, one should note that the probability distribution of $ \boldsymbol{Z}_1 $ conditioned on the parameter $ \theta $ should not be misspecified. Besides, the same as discussed in Theorem \ref{thm:prob}, the dimension of sufficient statistic $ \boldsymbol{T} $ should be the same as $ \boldsymbol{\theta} $, which was indicated by \citet[Section 2.5]{cox2006principles}.
\end{remark}

\subsection{Assumptions Conditional on Sufficient Statistics}\label{subsec3:assump}
Sufficient statistics play an important role in the aforementioned two theorems. To utilize them in our proposed method, we simply treat $ \{\boldsymbol{U}_i\}_{i=1}^m $ as cluster-specific parameters in Model \eqref{eq:glm_A_ij}, then we consider the following assumption for sufficient statistics existence.
\begin{assumption}\label{assum:sufficient}
	For each cluster, there exists a function of $ \boldsymbol{A}_i $, defined as $ \boldsymbol{T}_i =\boldsymbol{T}_i(\boldsymbol{A}_i) $,  is sufficient for $ \boldsymbol{U}_i $ in \eqref{eq:glm_A_ij}. Moreover, for any value $ \boldsymbol{t}$ of $ \boldsymbol{T}_i $ and any unit $ j $, there exist at least two different possible values of $ A_{ij} $ in $ \boldsymbol{A}_i $, i.e. $ a_{ij} $ in $ \boldsymbol{a}_i $ and $ a_{ij}^* $ in $ \boldsymbol{a}^*_i $, such that (i) $ a_{ij}\neq a_{ij}^* $ and (ii) $ \boldsymbol{T}_i(\boldsymbol{a}_i)=\boldsymbol{T}_i(\boldsymbol{a}_i^*)=\boldsymbol{t} $.
\end{assumption}  

Recall two aforementioned assumptions in Section \ref{sebsec2:assumptions}, Cluster-level Positivity (Assumption \ref{assum:positivity1}) and Cluster-level Ignorability (Assumption \ref{assum:ignorability1}). Both of them require the information of cluster-specific confounding variable $ \boldsymbol{U}_i $ in each cluster, which is not observed in data. Assume Assumption \ref{assum:sufficient} holds, by Theorems \ref{thm:prob} and \ref{thm:indep}, Assumptions \ref{assum:positivity1} and \ref{assum:ignorability1} can be replaced:

\renewcommand\theassumption{3*} 

\begin{assumption}
	\label{assum:positivity2} 
	The treatment assignment probability conditional on sufficient statistic satisfies $ 0< P\{{A}_{ij}={a}_{ij}|\boldsymbol{X}_i,\boldsymbol{T}_i=\boldsymbol{T}(\boldsymbol{a}_i)\}<1 $, for all $ i $, $ j $, and $ \boldsymbol{a}_i=(a_{i1},\ldots,a_{in_i}) $. When the treatment is binary, $ \boldsymbol{a}_i\neq \boldsymbol{0} $ or $ \boldsymbol{1} $.  
\end{assumption}

\renewcommand\theassumption{4*} 
\begin{assumption}
	\label{assum:ignorability2}
	$ \{\boldsymbol{Y}_{i}(0), \boldsymbol{Y}_{i}(1)\}$ $ \indept \boldsymbol{A}_{i} | \boldsymbol{X}_i, \boldsymbol{T}_i $ for all $ i $.
\end{assumption}

The general form of Assumption \ref{assum:ignorability2} is $ \{\boldsymbol{Y}_{i}(a)\}_{a\in\Omega_{A}} \indept \boldsymbol{A}_{i} | \boldsymbol{X}_i,\boldsymbol{T} _i $ for all $ i $. The above two assumptions are more preferable to the original Cluster-level Positivity and Cluster-level Ignorability in Assumptions \ref{assum:positivity1} and \ref{assum:ignorability1}. This is because, by utilizing the sufficient statistics $ \{\boldsymbol{T}_i\}_{i=1}^m $ in Theorems \ref{thm:prob} and \ref{thm:indep}, the unmeasured cluster-specific confounding variables $ \{\boldsymbol{U}_i\}_{i=1}^m $ can be ignored in Assumptions \ref{assum:positivity2} and \ref{assum:ignorability2}. Then methods proposed under these two assumptions can also be relaxed from considering $ \boldsymbol{U}_i $. 


\section{Proposed Methodology}\label{sec:prop.method}
\subsection{Inverse Conditional Probability Weighted (ICPW) estimator}\label{subsec3:cipw}
Our proposed estimator is constructed from a conditional probability by utilizing the sufficient statistic. In particular, based on model \eqref{eq:glm_A_ij}, we construct a probability of $ A_{ij} $ conditional on $ \boldsymbol{X}_{i} $ and the sufficient statistic $ \boldsymbol{T}_i $ described in  Assumption \ref{assum:sufficient}, 
\begin{align}\label{eq:cond_prob1}
P(A_{ij}=a_{ij}|\boldsymbol{X}_{i}, \boldsymbol{T}_i;\boldsymbol{\beta})
&=\frac{\sum_{\boldsymbol{a}^*\in\Omega_{i,j}}P(\boldsymbol{A}_{i}=\boldsymbol{a}_{i}^*|\boldsymbol{X}_{i},\boldsymbol{U}_i;\boldsymbol{\beta})}{\sum_{\tilde{\boldsymbol{a}}\in\tilde{\Omega}_{i}}P(\boldsymbol{A}_{i}=\tilde{\boldsymbol{a}}_{i}|\boldsymbol{X}_{i}, \boldsymbol{U}_i;\boldsymbol{\beta})}
\end{align}
for all $ i $, $ j $, and any value $ \boldsymbol{a}_i $ in the domain $ \Omega_i $. The set $ \Omega_{i,j} $ in the numerator of \eqref{eq:cond_prob1} is a set of all possible treatments $ \boldsymbol{a}^*=(a_1^*,\cdots,a_{n_i}^*) $ satisfying two criteria -- (i) the $ j $th components is the same as the observed value, i.e., $ a_j^*=a_{ij} $; (ii) the value of $ \boldsymbol{T}_i(\boldsymbol{a}^*) $ equals to the value of $ \boldsymbol{T}_i(\boldsymbol{a}_i) $ from data. In short, $ \Omega_{i,j} = \{\boldsymbol{a}^*\in \Omega_i|a_j^*=a_{ij} \text{ and }\boldsymbol{T}_i(\boldsymbol{a}^*)=\boldsymbol{T}_i(\boldsymbol{a}_i)\} $. The other set $ \tilde{\Omega}_i $ in the denominator of \eqref{eq:cond_prob1} is defined as $ \tilde{\Omega}_i = \{\tilde{\boldsymbol{a}}\in \Omega_i|\boldsymbol{T}_i(\tilde{\boldsymbol{a}})=\boldsymbol{T}_i(\boldsymbol{a}_i)\} $. In particular, $ \tilde{\Omega}_i $ contains all possible permutations of treatments within one cluster such that the function $ \boldsymbol{T}_i $ of each permutation is the same as that of the observed treatments in the cluster. For all units in the $ i $th cluster, we assign each unit a weight defined as the inverse of the conditional probability described in \eqref{eq:cond_prob1}. The conditional probability is an important component in the proposed method. So the inverse conditional probability weighted (ICPW) estimator for $ E\{Y(a)\} $ is
$
Y_{ICPW}(a)=\frac{1}{n}\sum_{i=1}^{m}\sum_{j=1}^{n_i}\frac{\mathcal{I}(A_{ij}=a)}{P(A_{ij}=a|\boldsymbol{X}_i,\boldsymbol{T}_i;\boldsymbol{\beta})}Y_{ij}. 
$

Instead of unit-level unbiasedness, we show that our proposed weighting method is cluster-level unbiased. That is, suppose $ E\{Y_{ij}(a)\} $ is finite for all $ i $, $ j $ and $ a $, then for a cluster-level potential outcome sum $ \sum_{j=1}^{n_i}E\{Y_{ij}(a)\} $ with treatment level $ a $,
\begin{align}\label{eq:unbias_cluster}
&E\Big\{\sum_{j=1}^{n_i}\frac{\mathcal{I}(A_{ij}=a)}{P(A_{ij}=a|\boldsymbol{X}_i,\boldsymbol{T}_i;\boldsymbol{\beta})}Y_{ij}\Big\}\nonumber\\
=&\sum_{j=1}^{n_i}E_{\boldsymbol{X}_{i},\boldsymbol{T}_i}\left[E_{A_{ij}, Y_{ij}(a)|\boldsymbol{X}_{i},\boldsymbol{T}_i}\Big\{\frac{\mathcal{I}(A_{ij}=a)}{P(A_{ij}=a|\boldsymbol{X}_{i},\boldsymbol{T}_i;\boldsymbol{\beta})}Y_{ij}(a)\bigg|\boldsymbol{X}_{i},\boldsymbol{T}_i \Big\}\right]\nonumber\\
=&\sum_{j=1}^{n_i}E_{\boldsymbol{X}_{i},\boldsymbol{T}_i}\left[E_{A_{ij}|\boldsymbol{X}_{i},\boldsymbol{T}_i}\Big\{\frac{\mathcal{I}(A_{ij}=a)}{P(A_{ij}=a|\boldsymbol{X}_{i}, \boldsymbol{T}_i;\boldsymbol{\beta})}\bigg|\boldsymbol{X}_i, \boldsymbol{T}_i \Big\} E_{Y_{ij}(a)|\boldsymbol{X}_{i},\boldsymbol{T}_i}\{Y_{ij}(a)|\boldsymbol{X}_{i},\boldsymbol{T}_i\} \right]\nonumber\\
=&\sum_{j=1}^{n_i}E_{Y_{ij}(a)}[Y_{ij}(a)]. 	
\end{align}
The above equation holds due to Assumptions \ref{assum:consistency1}, \ref{assum:ignorability2} and 
$
E\Big\{\frac{\mathcal{I}(A_{ij}=a)}{P(A_{ij}=a|\boldsymbol{X}_{i},\boldsymbol{T}_i;\boldsymbol{\beta})}\bigg|\boldsymbol{X}_{i},\boldsymbol{T}_i \Big\}
=1 $.

Therefore, for binary treatment, the corresponding ICPW estimator of the average treatment causal effect based on conditional probability described in \eqref{eq:cond_prob1} is
\begin{equation}\label{eq:icpw_tau} 
{\tau}_{ICPW}=\frac{1}{n}\sum_{i}^{m}\sum_{j}^{n_i}\Big\{\frac{A_{ij}Y_{ij}}{P(A_{ij}=1|\boldsymbol{X}_{i},T_i;\boldsymbol{\beta})} - \frac{(1-A_{ij})Y_{ij}}{1-P(A_{ij}=1|\boldsymbol{X}_{i},T_i;\boldsymbol{\beta})} \Big\}.
\end{equation}

For the aforementioned logistic model \eqref{eq:logistic_A_ij}, the sufficient statistic is the treatment sum in the cluster $ T_i=\sum_{j=1}^{n_i}A_{ij} $. Then the probability conditional on sufficient statistic is
\begin{equation}\label{eq:cp_logistic}
P({A}_{ij}={a}_{ij}|\boldsymbol{X}_i,\sum_{j=1}^{n_i}A_{ij}=\sum_{j=1}^{n_i}a_{ij};\boldsymbol{\beta})
=\frac{\exp(a_{ij}\boldsymbol{X}_{ij}^T{\boldsymbol{\beta}}) \sum_{\boldsymbol{a}^*\in \mathbb{V}_{i,j}} \exp(\sum_{l=1,l\ne j}^{n_i} a^*_{l}\boldsymbol{X}_{il}^T{\boldsymbol{\beta}})}{\sum_{\tilde{\boldsymbol{a}}\in \tilde{\mathbb{V}}_i}\exp(\sum_{l=1}^{n_i}\tilde{a}_{l}\boldsymbol{X}_{il}^T{\boldsymbol{\beta}})}
\end{equation}
for all $ i $, $ j $, and $ a_{ij}=0,1 $. Set $ \mathbb{V}_{i,j} $ in the numerator of \eqref{eq:cp_logistic} is a set of all possible treatments $ \boldsymbol{a}^*=(a_1^*,\cdots,a_{n_i}^*) $ satisfying two criteria -- (i) the $ j^th $ components is the same as the observed value, i.e. $ a_j^*=a_{ij} $; (ii) the component sum in $ \boldsymbol{a}^* $ equals to the unit treatment sum in one cluster in the dataset, i.e. $ \sum_{l=1}^{n_i}a_l^*=\sum_{l=1}^{n_i}a_{il} $. The definition notation of $ \mathbb{V}_{i,j} $ is $ \mathbb{V}_{i,j} = \{\boldsymbol{a}^*\in \{0,1\}^{n_i}|a^*_j=a_{ij}\text{ and }\sum_{l=1}^{n_i}a^*_{l}=\sum_{l=1}^{n_i}a_{il}\} $. The other set $ \tilde{\mathbb{V}}_i $ in the denominator of \eqref{eq:cp_logistic} has components $ \tilde{\boldsymbol{a}}=(\tilde{a}_{1},\ldots,\tilde{a}_{n_i}) $. The set is defined as $ \tilde{\mathbb{V}}_i = \{\tilde{\boldsymbol{a}}\in \{0,1\}^{n_i}|\sum_{l=1}^{n_i}\tilde{a}_{l}=\sum_{l=1}^{n_i}a_{il}\} $. Specifically, $ \tilde{\mathbb{V}}_i $ contains all possible permutations of treatments within a cluster such that the each permutation sum equals to the observed treatment sum in the cluster from data. Moreover, because $ \boldsymbol{a}_i\neq \boldsymbol{0},\boldsymbol{1} $, the conditional probabilities, $ P\{\boldsymbol{A}_i=\boldsymbol{0}|\boldsymbol{X}_i,\sum_{j=1}^{n_i}A_{ij}=0\}=P\{\boldsymbol{A}_i=\boldsymbol{1}|\boldsymbol{X}_i,\sum_{j=1}^{n_i}A_{ij}=n_i\}=1 $, are excluded in Assumption \ref{assum:ignorability2}. It is similar to the method proposed by \cite{skinner2011inverse}, where the nonresponse indicator is treated as a binary treatment.
The above ICPW method can be summarized in Algorithm \ref{algICPW}.

\begin{algorithm}[h]
	\caption{\label{algICPW} Inverse conditional probability weighted (ICPW) estimator for $ \tau $ with binary treatment}
	\begin{algorithmic}[1]
		\makeatletter
		\setcounter{ALG@line}{-1}
		\makeatother
		\State Let the treatment $ A_{ij} $ for unit $ j $ in cluster $ i $ follows model \eqref{eq:logistic_A_ij}, there exists a function of $ \boldsymbol{A}_i $, defined $ T_i $, satisfies Assumption \ref{assum:sufficient}.
		\State Obtain the conditional maximum likelihood estimator $ \hat{\boldsymbol{\beta}} $ by maximizing the joint conditional likelihood \eqref{eq:cond.like.a} with $ a_{ij}=0 $ or $ 1 $.
		\State Compute the conditional probability with the conditional maximum likelihood estimator $ \hat{\boldsymbol{\beta}} $. That is, compute $ P(A_{ij}=1|\boldsymbol{X}_i,T_i;\hat{\boldsymbol{\beta}}) $.
		\State Compute the ICPW estimator $ \hat{\tau}_{ICPW} $ for $ \tau $
		\begin{equation}\label{eq:tau_hat_icpw}
		\hat{\tau}_{ICPW}=\frac{1}{n}\sum_{i=1}^{m}\sum_{j=1}^{n_i}\Big\{\frac{A_{ij}Y_{ij}}{P(A_{ij}=1|\boldsymbol{X}_{i},T_i;\hat{\boldsymbol{\beta}})} - \frac{(1-A_{ij})Y_{ij}}{1-P(A_{ij}=1|\boldsymbol{X}_{i},T_i;\hat{\boldsymbol{\beta}})} \Big\}.
		\end{equation} 
	\end{algorithmic}
\end{algorithm}

For multiple treatments, e.g. the aforementioned multinomial logistic model \eqref{eq:multiple_trt}, the sufficient statistic for $\boldsymbol{U}_i$ is $\boldsymbol{T}_i=(T_{i0},\dots,T_{i(K-1)})$, where $T_{ik}=\sum_{j=1}^{n_i}\mathcal{I}(A_{ij}=k)$ for $k=0,\dots,K-1$. Then the conditional probability for $A_{ij}$ conditional on $\boldsymbol{T}_i$ is
\begin{align} 
P(A_{ij}=a_{ij}|\boldsymbol{X}_{ij},\boldsymbol{T}_{i})
= \frac{\exp\{\sum_{k=1}^{K}\mathcal{I}(a_{ij}=k)\boldsymbol{X}_{ij}^T\boldsymbol{\beta}_k+\mathcal{I}(a_{ij}=0)\}\lambda_{ij}}{\sum_{\tilde{\boldsymbol{a}}\in \tilde{\mathbb{V}}_{i}}\exp\left[\sum_{l=1}^{n_i}\{\sum_{k=1}^{K}\mathcal{I}(\tilde{a}_{l}=k)\boldsymbol{X}_{ij}^T\boldsymbol{\beta}_k+\mathcal{I}(\tilde{a}_{l}=0)\} \right]},
\end{align}
where $ \lambda_{ij}=\sum_{\boldsymbol{a}^*\in \mathbb{V}_{i,j}}\exp\left[\sum_{l=1,l\neq j}^{n_i}\{\sum_{k=1}^{K}\mathcal{I}(a_{l}^*=k)\boldsymbol{X}_{ij}^T\boldsymbol{\beta}_k+\mathcal{I}(a_{l}^*=0)\} \right] $. Set $ \mathbb{V}_{i,j} $ in the above equation is a set of all possible treatments $ \boldsymbol{a}^*=(a_1^*,\cdots,a_{n_i}^*) $ satisfying two criteria -- (i) the $ j $th components is the same as the observed value, i.e. $ a_j^*=a_{ij} $; (ii) the component sum in $ \boldsymbol{a}^* $ equals to the unit treatment sum in one cluster in the dataset for each treatment category in each cluster $\boldsymbol{T}_i(\boldsymbol{a}^*_i)=\boldsymbol{T}_i(\boldsymbol{a}_i)$, i.e. $ \sum_{l=1}^{n_i}\mathcal{I}(a_l^*=k)=\sum_{l=1}^{n_i}\mathcal{I}(a_{il}=k) $ for $k=1,\dots,K$. The definition notation of $ \mathbb{V}_{i,j} $ is $ \mathbb{V}_{i,j} = \{\boldsymbol{a}^*\in \{0,1,\dots,K\}^{n_i}|a^*_j=a_{ij}\text{ and }\sum_{l=1}^{n_i}\mathcal{I}(a_l^*=k)=\sum_{l=1}^{n_i}\mathcal{I}(a_{il}=k) \text{ for all }k\} $. The set $ \tilde{\mathbb{V}}_i $ has components $ \tilde{\boldsymbol{a}}=(\tilde{a}_{1},\ldots,\tilde{a}_{n_i}) $. It is defined as $ \tilde{\mathbb{V}}_i = \{\tilde{\boldsymbol{a}}\in \{0,1,\dots,K\}^{n_i}|\sum_{l=1}^{n_i}\mathcal{I}(\tilde{a}_{l}=k)=\sum_{l=1}^{n_i}\mathcal{I}(a_{il}=k)\text{ for all }k\} $. Specifically, $ \tilde{\mathbb{V}}_i $ contains all possible permutations of treatment $k$th category within a cluster such that the each permutation sum equals to the observed treatment sum in the cluster from data.
The algorithm for our proposed method is summarized in Algorithm \ref{algICPW1}. 

\begin{algorithm}[h]
	\caption{\label{algICPW1} Inverse conditional probability weighted (ICPW) estimator for $ E\{Y(a)\} $}
	\begin{algorithmic}[1]
		\makeatletter
		\setcounter{ALG@line}{-1}
		\makeatother
		\State Let the treatment $ A_{ij} $ for unit $ j $ in cluster $ i $ follows model \eqref{eq:glm_A_ij}, there exists a function of $ \boldsymbol{A}_i $, defined $ T_i $, satisfies Assumption \ref{assum:sufficient}.
		\State Obtain the conditional maximum likelihood estimator $ \hat{\boldsymbol{\beta}} $ by maximizing the joint conditional likelihood
		\begin{equation}\label{eq:cond.like.a}
		L^c(\boldsymbol{\beta}) =\prod_{i=1}^{m}\prod_{j=1}^{n_i}P(A_{ij}=a_{ij}|\boldsymbol{X}_i,T_i;\boldsymbol{\beta}),
		\end{equation}
		where $ P(A_{ij}=a_{ij}|\boldsymbol{X}_i,T_i;\boldsymbol{\beta}) $ is described in \eqref{eq:cond_prob1}.
		\State Compute the conditional probability with the conditional maximum likelihood estimator $ \hat{\boldsymbol{\beta}} $. That is, compute $ P(A_{ij}=a|\boldsymbol{X}_i,T_i;\hat{\boldsymbol{\beta}}) $.
		\State Compute the ICPW estimator $ \hat{Y}_{ICPW}(a) $ for $ E\{Y(a)\} $
		\begin{equation}\label{eq:ya_hat_icpw}
		\hat{Y}_{ICPW}(a)=\frac{1}{n}\sum_{i=1}^{m}\sum_{j=1}^{n_i}\frac{\mathcal{I}\{A_{ij}=a\}Y_{ij}}{P(A_{ij}=a|\boldsymbol{X}_{i},T_i;\hat{\boldsymbol{\beta}})} .
		\end{equation} 
	\end{algorithmic}
\end{algorithm}

\subsection{Robustness of ICPW estimator}
\begin{figure}[h]
	\centering
	\makebox{
		\begin{tikzpicture}[node distance = 1in]
		\node(X){$ \boldsymbol{X}_{i} $};
		\node[right of = X, yshift = -.0cm](A){$ A_{ij} $};
		\node[right of = A, yshift = -.0cm](Y){$ Y_{ij} $};
		\node[below of = A, yshift = 0.25in](U){$ U_i $};
		\draw [->, thick] (X) -- (A);
		\draw [->, thick] (A) -- (Y);
		\draw [->, thick] (X.north) to [out=45,in=135] (Y.north);	
		\draw [-, thick, loosely dashed] (U) -- node[left]{$\rho_{\boldsymbol{X},U}$} (X);
		\draw [->, thick] (U) -- (A);
		\draw [->, thick, loosely dashed] (U) -- node[right]{$\rho_{Y,U}$} (Y);
		\end{tikzpicture}
	}
	\caption{\label{fig:rho}A graph representing two possible correlations with respect to the unmeasured cluster-specific confounding variable $ \boldsymbol{U}_i $. One is the correlation ($\rho_{\boldsymbol{X},\boldsymbol{U}}$) between the covariates $ \boldsymbol{X}_{i} $ and $ \boldsymbol{U}_i $. The other one is the correlation ($\rho_{Y,\boldsymbol{U}}$) between the outcome $ Y_{ij} $ and $ \boldsymbol{U}_i $.}
\end{figure}
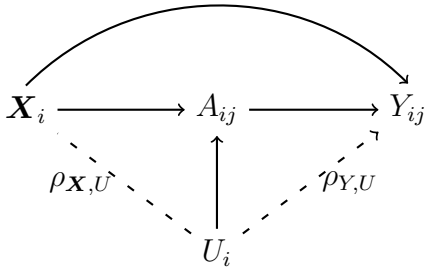

An attractive property of the ICPW estimator is its robustness, which is summarized below. 
\begin{theorem}\label{thm:icpw_robust}
	The proposed ICPW estimator is robust to both (i) the correlation between the unmeasured cluster-specific confounding variable and the covariates and (ii) the correlation between the unmeasured cluster-specific confounding variable and the outcome. 
\end{theorem}
The proof is in the Supplementary Materials.
\begin{remark}
	Theorem \ref{thm:icpw_robust} illustrates the unbiasedness holds no matter the correlation between $ \boldsymbol{X}_i $ and $ \boldsymbol{U}_i $ (i.e., $ \rho_{\boldsymbol{X},{U}} $ in Figure \ref{fig:rho}), or the correlation between $ \boldsymbol{Y}_i $ and $ \boldsymbol{U}_i $ (i.e., $ \rho_{Y,\boldsymbol{U}} $ in Figure \ref{fig:rho}), or the characteristics of $ \boldsymbol{U}_i $ (e.g. its distribution) is. Since $ \boldsymbol{U}_i $s are not observed in the real data, the two correlations $ \rho_{\boldsymbol{X},\boldsymbol{U}} $ and $ \rho_{Y,\boldsymbol{U}} $ in Figure \ref{fig:rho} are usually unobserved too. Such robust property exhibits an advantage of ICPW method in that it comes with more flexibility and confidence in estimating the average causal effect. 
\end{remark}

\section{Asymptotic Properties}\label{sec:asymp}
The main goal of this section is to show the asymptotic properties of the ICPW estimator as $ n\rightarrow\infty $. To reach this goal, we first focus on the asymptotic properties of conditional maximum likelihood estimator (CMLE) of $ \boldsymbol{\beta} $. To make sure the CMLE of $ \boldsymbol{\beta} $ is uniquely determined, we consider the minimal sufficient statistic $ \boldsymbol{T}_i $ for $ \boldsymbol{U}_i $ for all $ i $. This is because, as stated in \cite{andersen1970asymptotic}, the conditional probability has less information about $ \boldsymbol{U}_i $ if $ \boldsymbol{T}_i $ is not minimum sufficient. Next, we prove the asymptotic properties of the ICPW estimator of $ E[Y(a)] $ for all treatment level $ a\in\Omega_{A} $. Lastly, the asymptotic properties of the ICPW estimator of $ \tau $ with binary treatment can be proved by Delta method using Taylor series expansion. Here we consider the asymptotic results with respect to the number of clusters. That is, we will investigate the asymptotic properties of ICPW estimator with respect to $ m $ when $ n_i $'s are fixed and bounded. 

\subsection{Asymptotic Properties of CMLE for $ \boldsymbol{\beta} $}
\cite{andersen1970asymptotic} proved that the conditional maximum likelihood estimates are consistent and asymptotically normally distributed under regularity conditions. We adopt \cite{andersen1970asymptotic}'s results to show the asymptotic properties of CMLE for $ \beta $.  

\begin{theorem}(Consistency of CMLE for $ \boldsymbol{\beta} $)\label{thm:cons.beta}
	Suppose that Assumption 1 and the Conditions 1-3 specified in the Web Appendix A hold, and the treatment assignment follows the cluster-specific model in \eqref{eq:glm_A_ij}, and there exist sufficient statistics $  \boldsymbol{T}_i $ as specified in Assumption 5 and $ \max_{1\leq i\leq m}n_i/n\rightarrow 0 $ as $ n\rightarrow\infty $. The CMLE $ \hat{\boldsymbol{\beta}} $ can be obtained by maximizing the joint conditional likelihood $ \prod_{i=1}^{m}\prod_{j=1}^{n_i} P(A_{ij}=a_{ij}|\boldsymbol{X}_{i},  \boldsymbol{T}_i;\boldsymbol{\beta}) $, where $ P(A_{ij}=a_{ij}|\boldsymbol{X}_{i},  \boldsymbol{T}_i;\boldsymbol{\beta}) $ is specified in \eqref{eq:cond_prob1}. Therefore, $ \hat{\boldsymbol{\beta}} $ is a consistent estimate for $ \boldsymbol{\beta} $.
\end{theorem}

\begin{theorem}(Asymptotic Normality of CMLE for $ \boldsymbol{\beta} $)\label{thm:assym.beta}
	Suppose that Assumption 1 and the Conditions 1-5 specified in the Web Appendix A hold, and the treatment assignment follows the cluster-specific model in \eqref{eq:glm_A_ij}, and there exist sufficient statistics $  \boldsymbol{T}_i $ for all $ i $ as specified in Assumption 5 and $ \max_{1\leq i\leq m}n_i/n\rightarrow 0 $ as $ n\rightarrow\infty $. The CMLE $ \hat{\boldsymbol{\beta}} $ can be obtained by maximizing the joint conditional likelihood $ \prod_{i=1}^{m}\prod_{j=1}^{n_i} P(A_{ij}=a_{ij}|\boldsymbol{X}_{i},  \boldsymbol{T}_i;\boldsymbol{\beta}) $, where $ P(A_{ij}=a_{ij}|\boldsymbol{X}_{i},  \boldsymbol{T}_i;\boldsymbol{\beta}) $ is specified in \eqref{eq:cond_prob1}. Let $ \phi_{ij}(a|\boldsymbol{X}_i,  \boldsymbol{T}_i; \boldsymbol{\beta}) $ represent the conditional probability density function for $ P(A_{ij}=a|\boldsymbol{X}_{i},  \boldsymbol{T}_i;\boldsymbol{\beta}) $, which is continuous and differentiable with respect to $ \boldsymbol{\beta} $ at $ \boldsymbol{\beta}_0 $. Then we have
	$
	\sqrt{n}(\hat{\boldsymbol{\beta}} - \boldsymbol{\beta}_0) \rightarrow N(\boldsymbol{0}, B_1(\boldsymbol{\beta}_0)) 
	$
	in distribution as $ n\rightarrow\infty $, where $ B_1(\boldsymbol{\beta})=\{B_2(\boldsymbol{\beta})\}^{-1}B_3(\boldsymbol{\beta})\{B_2(\boldsymbol{\beta})\}^{-1}  $ with 
	\begin{equation*}
	B_2(\boldsymbol{\beta})=\frac{1}{n}\sum_{i=1}^{m}\sum_{j=1}^{n_i} E\Big\{\frac{\partial}{\partial\boldsymbol{\beta}\partial\boldsymbol{\beta}^T}\log\phi_{ij}(A_{ij}|\boldsymbol{X}_i, \boldsymbol{T}_i,\boldsymbol{\beta})\Big\}
	\end{equation*}
	and 
	\begin{equation*}
	B_3(\boldsymbol{\beta})=\frac{1}{n}\sum_{i=1}^{m} E\left[\Big\{\sum_{j=1}^{n_i}\frac{\partial}{\partial\boldsymbol{\beta}}\log\phi_{ij}(A_{ij}|\boldsymbol{X}_i, \boldsymbol{T}_i,\boldsymbol{\beta})\Big\}\Big\{\sum_{j=1}^{n_i}\frac{\partial}{\partial\boldsymbol{\beta}^T}\log\phi_{ij}(A_{ij}|\boldsymbol{X}_i, \boldsymbol{T}_i,\boldsymbol{\beta})\Big\}\right].
	\end{equation*}	
\end{theorem}

The proofs of Theorems \ref{thm:cons.beta} and \ref{thm:assym.beta} are skipped here since they are conceptually the same as \cite{andersen1970asymptotic}'s proof. The difference is that Andersen's work was not aimed to clustered data structure. To adopt his work to clustered data, we treat the cluster-level joint probability $ \prod_{j=1}^{n_i} \phi_{ij}(A_{ij}|\boldsymbol{X}_{i},  \boldsymbol{T}_i;\boldsymbol{\beta}) $ as the unit probability in his proof. Therefore, the consistency result can be obtained with respect to the number of clusters (i.e. $ n \rightarrow \infty $ as $ m\rightarrow\infty $). 

\subsection{Asymptotic Properties of ICPW estimator for $ E[Y(a)] $} \label{subsec:asym.Ya}
In Section \ref{subsec3:cipw}, we have shown in \eqref{eq:unbias_cluster} that the proposed ICPW estimator is an  unbiased estimator for one cluster. Therefore, it is straight forward to show the overall unbiasedness:
\begin{equation}\label{eq:consis.ya}
E\Big\{\frac{1}{n}\sum_{i}^{m}\sum_{j}^{n_i}\frac{\mathcal{I}\{A_{ij}=a\}Y_{ij}}{P(A_{ij}=a|\boldsymbol{X}_{i}, \boldsymbol{T}_i;\boldsymbol{\beta})}\Big\} = \frac{1}{n}\sum_{i}^{m}\sum_{j}^{n_i} E\{ Y_{ij}(a) \} = E\{Y(a)\} ,\ \forall a\in\Omega_{A} .
\end{equation}
The corresponding estimator for $ E\{Y(a)\} $, i.e., $ \hat{Y}_{ICPW}(a) $, is defined in \eqref{eq:ya_hat_icpw} in Algorithm 2. The asymptotic properties of $ \hat{Y}_n(a) $ is shown in Theorem \ref{thm:assym.ya}.

\begin{theorem}(Asymptotic Normality of ICPW estimator for $ E[Y(a)] $)\label{thm:assym.ya}
	Suppose $ \sqrt{n}(\hat{\boldsymbol{\beta}} - \boldsymbol{\beta}_0) \rightarrow N(\boldsymbol{0}, B_1(\boldsymbol{\beta}_0)) $ and $ \phi_{ij}(a|\boldsymbol{X}_i,  \boldsymbol{T}_i; \boldsymbol{\beta}) $ is continuous and differentiable with respect to $ \boldsymbol{\beta} $ at $ \boldsymbol{\beta}_0 $, with $ \frac{\partial\phi_{ij}(a|\boldsymbol{X}_i,  \boldsymbol{T}_i; \boldsymbol{\beta})}{\partial\boldsymbol{\beta}} |_{\boldsymbol{\beta}=\boldsymbol{\beta}_0}\neq \boldsymbol{0} $. Let $ \hat{Y}_n(\boldsymbol{\beta}_0;a) $ be the ICPW estimator at $ \boldsymbol{\beta}=\boldsymbol{\beta}_0 $, and $ E\{Y(a)\} $ be the expectation of the ICPW estimator at $ \boldsymbol{\beta}=\boldsymbol{\beta}_0 $. Assume $ \sigma^2_{1,\boldsymbol{\beta}_0} = E[\hat{Y}_n(\boldsymbol{\beta}_0;a) - E\{Y(a)\} ]^2 $ is bounded. Then the ICPW estimator in \eqref{eq:ya_hat_icpw} satisfies
	$
	\sqrt{n}V_1(\boldsymbol{\beta}_0)^{-1/2}(\hat{Y}_{ICPW}(a) - E\{Y(a)\}) \rightarrow N({0}, 1) 
	$
	in distribution as $ n\rightarrow\infty $, where $ V_1(\boldsymbol{\beta}_0) = E\{ H_1(\boldsymbol{\beta}_0)^T B_1(\boldsymbol{\beta}_0) H_1(\boldsymbol{\beta}_0) \}$ is assumed to be bounded and positive, and  
	$
	H_1(\boldsymbol{\beta}) = \frac{1}{n}\sum_{i=1}^{m}\sum_{j=1}^{n_i}\frac{\mathcal{I}\{A_{ij}=a\}Y_{ij}}{\phi_{ij}^2(a|\boldsymbol{X}_i,  \boldsymbol{T}_i; \boldsymbol{\beta})}\frac{\partial\phi_{ij}(a|\boldsymbol{X}_i,  \boldsymbol{T}_i; \boldsymbol{\beta})}{\partial\boldsymbol{\beta}}.
	$
\end{theorem}
The proof is in the Supplementary Materials.

\subsection{Asymptotic Properties of ICPW estimator for $ \tau $}
Form the results in \eqref{eq:consis.ya} in Section \ref{subsec:asym.Ya}, we know the ICPW estimator for binary treatment is unbiased for $ \tau $. The asymptotic properties of $ \hat{\tau}_{ICPW} $ defined in \eqref{eq:tau_hat_icpw} is presented below:

\begin{theorem}(Asymptotic Normality of ICPW estimator for $ \tau $)\label{thm:assym.tau}
	For binary treatment, let $ \phi_{ij}(\boldsymbol{\beta}) $ represent the conditional probability density function of $ P(A_{ij}=1|\boldsymbol{X}_{i},  \boldsymbol{T}_i;\boldsymbol{\beta}) $. Suppose $ \sqrt{n}(\hat{\boldsymbol{\beta}} - \boldsymbol{\beta}_0) \rightarrow N(\boldsymbol{0}, B_1(\boldsymbol{\beta}_0)) $ and $ \phi_{ij}(\boldsymbol{\beta}) $ is  continuous and differentiable with respect to $ \boldsymbol{\beta} $ at $ \boldsymbol{\beta}_0 $, with $ \frac{\partial\phi_{ij}(\boldsymbol{\beta})}{\partial\boldsymbol{\beta}} |_{\boldsymbol{\beta}=\boldsymbol{\beta}_0}\neq \boldsymbol{0} $. $ \tau $ is the true average causal effect at $ \boldsymbol{\beta}=\boldsymbol{\beta}_0 $, and let $ \hat{\tau}_n(\boldsymbol{\beta}_0) $ be the ICPW estimator at $ \boldsymbol{\beta}=\boldsymbol{\beta}_0 $. Assume $ \sigma^2_{2,\boldsymbol{\beta}_0} = E[\hat{\tau}_n(\boldsymbol{\beta}_0) - \tau ]^2 $ is bounded. Then the ICPW estimator in \eqref{eq:tau_hat_icpw} satisfies
	$
	\sqrt{n} V_2(\boldsymbol{\beta}_0)^{-1/2}(\hat{\tau}_{ICPW} - \tau) \rightarrow N(0,1) 
	$
	in distribution as $ n\rightarrow\infty $, where $ 	V_2(\boldsymbol{\beta}_0) = E\{ H_2(\boldsymbol{\beta}_0)^T B_1(\boldsymbol{\beta}_0) H_2(\boldsymbol{\beta}_0) \} $, which is assumed to be bounded and positive, and 
	$
	H_2(\boldsymbol{\beta}) =  \frac{1}{n}\sum_{i=1}^{m}\sum_{j=1}^{n_i}\Big\{\frac{A_{ij}}{\phi_{ij}^2(\boldsymbol{\beta})}+\frac{1-A_{ij}}{[1-\phi_{ij}(\boldsymbol{\beta})]^2}\Big\}Y_{ij}\frac{\partial\phi_{ij}(\boldsymbol{\beta})}{\partial\boldsymbol{\beta}}.
	$
\end{theorem}
The proof is in the Supplementary Materials.

\section{Simulation Studies}\label{sec:simu}
We conduct two simulation studies to show the robustness of the ICPW estimator. In the first simulation study, we specify the number of clusters to be $ m=500 $, and the cluster size ($ n_i $) to be the integer part of $ D_i\sim Unif(2,6) $. So cluster sizes range from 2 to 5. In comparison, the second simulation study has smaller data size. There are with 20 clusters ($ m=20 $) and the cluster size is the integer part of $ D_i\sim Unif(2,21) $, indicating a range from 2 to 20. Two covariates, a continuous covariate $ X_{1,ij} $ and a categorical covariate $ X_{2,ij} $, are generated independently for each unit. In particular, $ X_{1,ij}\sim N(0,1) $ and $ X_{2,ij}=-1, $ or $ 0 $, or $ 1 $ with equal probabilities. The cluster-specific confounding variable $ U_i\sim N(-\rho_{\boldsymbol{X},U}[\bar{X}_{1,i\cdot}+\bar{X}_{2,i\cdot}], 1) $, where $ \bar{X}_{1,i\cdot} $ and $ \bar{X}_{2,i\cdot} $ are the means over the units within one cluster. We change the value of $ \rho_{\boldsymbol{X},U} $ to manipulate the correlation between the covariates and $U_i$'s. Note that the expectation of $ U_i $ is always 0 for any $ \rho_{\boldsymbol{X},U} $. The treatment assignment mechanism is $ P(A_{ij}=1|\boldsymbol{X}_i,U_i)=\exp(X_{1,ij}+X_{2,ij}+U_i)/\{1+\exp(X_{1,ij}+X_{2,ij}+U_i)\} $.  
For each unit, two potential outcomes are generated as $ Y_{ij}(0)=X_{1,ij}+X_{2,ij}+e_{ij}^0 $ and $ Y_{ij}(1)=X_{1,ij}+X_{2,ij}+ \tau + \rho_{Y,U}U_i + e_{ij}^1 $, where $ \tau=2 $, $ e_{ij}^0, e_{ij}^1\sim N(0,1) $, and $ \rho_{Y,U} $ controls the correlation between the causal effect and $U_i$'s. The observed outcomes follow Assumption \ref{assum:consistency1}. We consider four scenarios:
\begin{enumerate}
	\item ($ \rho_{\boldsymbol{X},U}, \rho_{Y,U} $)=(0,0). The cluster-specific confounding variable $ U_i $ is independent of both the covariate $ \boldsymbol{X}_{i} $ and the causal effect $ Y_{ij}(1)-Y_{ij}(0) $;
	\item ($ \rho_{\boldsymbol{X},U}, \rho_{Y,U} $)=(5,0). The cluster-specific confounding variable $ U_i $ is correlated of the covariate $ \boldsymbol{X}_{i} $, and it is independent with the causal effect $ Y_{ij}(1)-Y_{ij}(0) $;
	\item ($ \rho_{\boldsymbol{X},U}, \rho_{Y,U} $)=(0,5). The cluster-specific confounding variable $ U_i $ is correlated of the causal effect $ Y_{ij}(1)-Y_{ij}(0) $, and it is independent with the covariate $ \boldsymbol{X}_{i} $;
	\item ($ \rho_{\boldsymbol{X},U}, \rho_{Y,U} $)=(5,5). The cluster-specific confounding variable $ U_i $ is correlated with both the covariate $ \boldsymbol{X}_{i} $ and the causal effect $ Y_{ij}(1)-Y_{ij}(0) $. 
\end{enumerate}
We obtain an estimator from each simulated data, i.e., $ \hat{\tau}_{simu}=1/n\sum_{i=1}^{m}\sum_{j=1}^{n_i}\{Y_{ij}(1)-Y_{ij}(0)\} $. Note that $ \hat{\tau}_{simu} $ can not be obtained from the real data due to fundamental problem in causal inference \citep{rubin1974estimating, holland1986statistics}.  Therefore $ \hat{\tau}_{simu} $ and $ se(\hat{\tau}_{simu}) $ are not used for comparison to other methods, but for an illustration of the true causal effect and its corresponding standard error obtained from simulated data. 

For method comparison, we consider four estimators for $ \tau $. The first is $ \hat{\tau}_{naive} $, which is a simple estimator without weight adjustment, i.e., $ \hat{\tau}_{naive} = 1/n\{A_{ij}Y_{ij} - (1-A_{ij})Y_{ij} \} $. The second estimator $ \hat{\tau}_{IPW,ran} $ is an IPW estimator in \eqref{eq:ipw_tau} by specifying \eqref{eq:glm_A_ij} as a logistic mixed effects model where cluster-specific effect is random. The third estimator $ \hat{\tau}_{IPW,fix} $ is an IPW estimator in \eqref{eq:ipw_tau} by specifying \eqref{eq:glm_A_ij} as a logistic model where cluster-specific effect is fixed effect. The last estimator $ \hat{\tau}_{ICPW} $ is the proposed estimator obtained from Algorithm \ref{algICPW}.

Simulation results are presented in Tables \ref{tab_simu1} and \ref{tab_simu2}. Each simulation study is conducted in R and are repeated 1,000 times. The simple estimator $ \hat{\tau}_{naive} $ shows large bias in general.
The IPW estimator $ \hat{\tau}_{IPW,ran} $ is biased when $ U_i $ is correlated with either covariates or the causal effect. Its bias becomes the largest in scenario 4. In comparison, the bias of $ \hat{\tau}_{IPW,fix} $ is not that large in both simulation studies. But $ \hat{\tau}_{IPW,fix} $ has the largest variance across all scenarios. This is resulted from the Neyman-Scott incidental parameter problem \citep{neyman1948consistent}. In particular, the variance of $ \hat{\tau}_{IPW,fix} $  is increased by the involvement of the cluster-specific parameters. Our proposed estimator $ \hat{\tau}_{ICPW} $ works well across all scenarios in both simulation studies. This confirms Theorem \ref{thm:icpw_robust} that the ICPW estimator is robust when cluster-specific confounding variable is correlated with the covariates and/or the causal effect.

\begin{table}
	\caption{\label{tab_simu1}Results of simulation study 1 based on 1,000 repetitions. Each repetition contains 500 clusters and the cluster size range from 2 to 5. The expected average causal effect is $ \tau=2 $. For each method, the estimate, bias to the expected average causal effect, and standard error (s.e.) are reported. }
	\centering  
	\begin{tabular}{|cccc|cccc|}
		\hline
		Method & Estimate & Bias to $ \tau $ & s.e. & Method & Estimate & Bias to $ \tau $ & s.e. \\ \hline
		\multicolumn{4}{|c|}{Scenario 1: ($ \rho_{\boldsymbol{X},U}, \rho_{Y,U} $)=(0,0)} &			\multicolumn{4}{c|}{Scenario 2: ($ \rho_{\boldsymbol{X},U}, \rho_{Y,U} $)=(5,0)}\\
		$ \hat{\tau}_{simu} $ & 2.000 & 0.000 & 0.034 & $ \hat{\tau}_{simu} $ & 2.001 & 0.001 & 0.034 \\
		$ \hat{\tau}_{naive} $ & 1.594 & -0.406 & 0.040 & $ \hat{\tau}_{naive} $ & 1.370 & -0.630 & 0.042  \\
		$ \hat{\tau}_{IPW,ran} $ & 2.009 & 0.009 & 0.072 & $ \hat{\tau}_{IPW,ran} $ & 1.914 & -0.086 & 0.063 \\
		$ \hat{\tau}_{IPW,fix} $ & 1.843 & -0.157 & 0.562 &	$ \hat{\tau}_{IPW,fix} $ & 1.906 & -0.094 & 0.522 \\		
		$ \hat{\tau}_{ICPW} $ & 2.003 & 0.003 & 0.148 & $ \hat{\tau}_{ICPW} $ & 2.003 & 0.003 & 0.137 \\	\hline
		\multicolumn{4}{|c|}{Scenario 3: ($ \rho_{\boldsymbol{X},U}, \rho_{Y,U} $)=(0,5)} & 			\multicolumn{4}{c|}{Scenario 4: ($ \rho_{\boldsymbol{X},U}, \rho_{Y,U} $)=(5,5)}\\
		$ \hat{\tau}_{simu} $ & 2.000 & 0.000 & 0.236 & $ \hat{\tau}_{simu} $ & 2.052 & 0.052 & 0.664 \\
		$ \hat{\tau}_{naive} $ & 2.022 & 0.022 & 0.140 & $ \hat{\tau}_{naive} $ & 3.414 & 1.414 & 0.361 \\
		$ \hat{\tau}_{IPW,ran} $ & 3.100 & 1.100 & 0.285 & $ \hat{\tau}_{IPW,ran} $ & 8.746 & 6.746 & 0.669 \\
		$ \hat{\tau}_{IPW,fix} $ & 1.619 & -0.381 & 1.505 & $ \hat{\tau}_{IPW,fix} $ & 0.996 & -1.004 & 2.913 \\		
		$ \hat{\tau}_{ICPW} $ & 2.005 & 0.005 & 0.400 & $ \hat{\tau}_{ICPW} $ & 2.089 & 0.089 & 1.016 \\	\hline
	\end{tabular}
\end{table}

	\begin{table}
	\caption{\label{tab_simu2}Results of simulation study 2 based on 1,000 repetitions. Each repetition contains 20 clusters and the cluster size range from 2 to 20. The expected average causal effect is $ \tau=2 $. For each method, the estimate, bias to the expected average causal effect, and standard error (s.e.) are reported. }
	\centering  
	\begin{tabular}{|cccc|cccc|}
		\hline
		Method & Estimate & Bias to $ \tau $ & s.e. & Method & Estimate & Bias to $ \tau $ & s.e. \\ \hline
		\multicolumn{4}{|c|}{Scenario 1: ($ \rho_{\boldsymbol{X},U}, \rho_{Y,U} $)=(0,0)} &			\multicolumn{4}{c|}{Scenario 2: ($ \rho_{\boldsymbol{X},U}, \rho_{Y,U} $)=(5,0)}\\
		$ \hat{\tau}_{simu} $ & 2.001 & 0.001 & 0.097 & $ \hat{\tau}_{simu} $ & 2.006 & 0.006 & 0.097 \\
		$ \hat{\tau}_{naive} $ & 1.578 & -0.422 & 0.144 & $ \hat{\tau}_{naive} $ & 1.375 & -0.625 & 0.163 \\
		$ \hat{\tau}_{IPW,ran} $ & 1.885 & -0.115 & 0.288 & $ \hat{\tau}_{IPW,ran} $ & 1.405 & -0.595 & 0.485 \\
		$ \hat{\tau}_{IPW,fix} $ & 1.981 & -0.019 & 0.438 & $ \hat{\tau}_{IPW,fix} $ & 1.973 & -0.027 & 0.618 \\		
		$ \hat{\tau}_{ICPW} $ & 2.010 & 0.010 & 0.366 & $ \hat{\tau}_{ICPW} $ & 2.007 & 0.007 & 0.442 \\	\hline
		\multicolumn{4}{|c|}{Scenario 3: ($ \rho_{\boldsymbol{X},U}, \rho_{Y,U} $)=(0,5)} & 			\multicolumn{4}{c|}{Scenario 4: ($ \rho_{\boldsymbol{X},U}, \rho_{Y,U} $)=(5,5)}\\
		$ \hat{\tau}_{simu} $ & 1.943 & 0.057 & 1.261 & $ \hat{\tau}_{simu} $ & 2.041 & 0.041 & 2.380 \\
		$ \hat{\tau}_{naive} $ & 2.280 & 0.280 & 0.761 & $ \hat{\tau}_{naive} $ & 3.504 & 1.504 & 1.404 \\
		$ \hat{\tau}_{IPW,ran} $ & 3.745 & 1.745 & 1.469 & $ \hat{\tau}_{IPW,ran} $ & 9.150 & 7.150 & 3.440 \\
		$ \hat{\tau}_{IPW,fix} $ & 1.931 & -0.069 & 1.494 & $ \hat{\tau}_{IPW,fix} $ & 1.850 & -0.150 & 4.827 \\		
		$ \hat{\tau}_{ICPW} $ & 1.989 & -0.011 & 1.430 & $ \hat{\tau}_{ICPW} $ & 2.029 & 0.029 & 3.423 \\	\hline
	\end{tabular}
\end{table}


\section{A Case Study}\label{sec:case}
For real data analysis, we apply the ICPW method to the low birth weight data from \citet{hosmer2000applied}. The data was collected from 189 women in 1986. Among these women, 59 had low-birth-weight babies and 130 had normal-weight babies. They were grouped according to their age. We are interested in estimating the average causal effect ($ \tau $) of mother smoking behavior ($ A=1 $ if yes and 0 if no smoking) to the baby birth weight in grams ($ Y $) among these women. After excluding clusters that violate the Assumption \ref{assum:positivity1}, we have 182 women in 20 clusters ($ m=20 $). In each cluster, there are 2 to 18 women ($ n_i $ ranges from 2 to 18). The covariates include race ($ X_1 $: white, black, and other), number of false premature labors ($ X_2 $), and standardized mother's weight at last menstrual period ($ X_3 $). 

Similar to the simulation studies, four methods are considered here: (i) the simple estimator, $ \hat{\tau}_{naive} $, without any weight adjustment; (ii) $ \hat{\tau}_{IPW,ran} $, the IPW estimator by fitting a logistic mixed effects model to the treatment, where the linear predictors include all three covariates and the cluster-specific effect is random; (iii) $ \hat{\tau}_{IPW,fix} $, the IPW estimator by fitting a logistic model similar to the model in (ii) except the cluster-specific effect is fixed effect; (iv) $ \hat{\tau}_{ICPW} $, the proposed ICPW method, where the linear predictors include all three covariates.   

Results with 100 bootstrap replicates are displayed in Table \ref{tab_case}. Among all estimates for the average causal effect, three estimates, except $ \hat{\tau}_{IPW,ran} $, are negative. The negative causal effect estimate indicates that mother smoking behavior reduces baby birth weight. The ICPW method presents a negative causal effect estimate. The corresponding $ 95\% $ confidence interval includes zero, indicating non-significant causal effect among these women. 

Moreover, we find some similarities by comparing this case study to the scenario 4 of simulation study 2 in the previous section. First, the number of clusters ($ m $) and cluster size range in both real data and simulated data are very close. Second, estimator $ \hat{\tau}_{IPW,ran} $ shows great difference to both $ \hat{\tau}_{IPW,fix} $ and $ \hat{\tau}_{ICPW} $. Third, the standard error of $ \hat{\tau}_{IPW,fix} $ is greater than that of $ \hat{\tau}_{ICPW} $. From these similarities, our conjecture is that the unmeasured cluster-specific confounding factors may be correlated with the covariates and the causal effect as the setting in scenario 4. This seems reasonable in this study that mother's age may be correlated with mother's covariates and baby's birth weight.

\begin{table}[h]
	\caption{\label{tab_case}Results of case study based on 100 bootstrap replicates. For each method, the estimate, standard error (s.e.), and 95\% confidence interval (c.i.) are reported. }
	\centering
	\fbox{%
		\begin{tabular}{crrc}
			Method & Estimate & s.e. & 95\% c.i. \\ \hline
			$ \hat{\tau}_{naive} $ & -705.9 & 46.6 & (-797.7, -628.9) \\
			$ \hat{\tau}_{IPW,ran} $ & 194.2 & 1353.3 & (-1445.9, 4787.0) \\
			$ \hat{\tau}_{IPW,fix} $ & -283.4 & 1898.7 & (-2985.0, 2864.2) \\	
			$ \hat{\tau}_{ICPW} $ & -227.6 & 402.3 & (-1108.7, 429.7) \\	
	\end{tabular}}
\end{table}

\section{Discussion}\label{sec:discuss}
The ICPW method is attractive for two reasons. First, it is robust to both correlation between 
$ U_i $ and the covariates, and the correlation between $ U_i $ and the outcome. Since $ U_i $ is unmeasured in data, it is usually difficult or impossible to obtain its correlations to other measured variables. Such correlations can result biased causal effect estimates in many methods. Comparatively, the robustness of ICPW method can overcome the unclear correlations. Second, we do not have to make any further assumptions on $ U_i $. Such assumptions include assuming $ U_i $ is a random effect, or is a fixed effect, or follows a prior distribution.  The relaxedness of further assumptions on $ U_i $ makes it more adaptable in estimation. 

Besides, it should be noted that our focus in this paper is the theoretical study of the ICPW method. In additional to the appealing theoretical properties, there are still some future work on the ICPW method that are worth exploring. 
First, when the cluster size is large, the computational load for implementing the ICPW method might increase. In particular, we need to consider all possible permutations in both the numerator and denominator of \eqref{eq:cond_prob1}. It will be a topic for future research to design numerical algorithms for computing the ICPW estimator efficiently under large cluster sizes. Second, the proposed ICPW method is originated from the simplest format of the IPW method. There are opportunities to make modifications to the ICPW method under more complex settings, for example time-varying treatment causal effect estimation.
	
\section*{Acknowledgments}
The author would like to thank Dr. Chong Wang for his helpful comments to the improvement of the paper.

\bibliographystyle{chicago}
\bibliography{ICPW_reference}	

\begin{thebibliography}{}

\bibitem[\protect\citeauthoryear{Andersen}{Andersen}{1970}]{andersen1970asymptotic}
Andersen, E.~B. (1970).
\newblock Asymptotic properties of conditional maximum-likelihood estimators.
\newblock {\em Journal of the Royal Statistical Society, Series B\/}, 283--301.

\bibitem[\protect\citeauthoryear{Arpino and Cannas}{Arpino and
  Cannas}{2016}]{arpino2016propensity}
Arpino, B. and M.~Cannas (2016).
\newblock Propensity score matching with clustered data. an application to the
  estimation of the impact of caesarean section on the apgar score.
\newblock {\em Statistics in Medicine\/}~{\em 35\/}(12), 2074--2091.

\bibitem[\protect\citeauthoryear{Arpino and Mealli}{Arpino and
  Mealli}{2011}]{arpino2011specification}
Arpino, B. and F.~Mealli (2011).
\newblock The specification of the propensity score in multilevel observational
  studies.
\newblock {\em Computational Statistics \& Data Analysis\/}~{\em 55\/}(4),
  1770--1780.

\bibitem[\protect\citeauthoryear{Austin and Stuart}{Austin and
  Stuart}{2015}]{austin2015moving}
Austin, P.~C. and E.~A. Stuart (2015).
\newblock Moving towards best practice when using inverse probability of
  treatment weighting (iptw) using the propensity score to estimate causal
  treatment effects in observational studies.
\newblock {\em Statistics in Medicine\/}~{\em 34\/}(28), 3661--3679.

\bibitem[\protect\citeauthoryear{Brumback and He}{Brumback and
  He}{2011}]{brumback2011adjusting}
Brumback, B.~A. and Z.~He (2011).
\newblock Adjusting for confounding by neighborhood using complex survey data.
\newblock {\em Statistics in Medicine\/}~{\em 30\/}(9), 965--972.

\bibitem[\protect\citeauthoryear{Brumback, Hern{\'a}n, Haneuse, and
  Robins}{Brumback et~al.}{2004}]{brumback2004sensitivity}
Brumback, B.~A., M.~A. Hern{\'a}n, S.~J. Haneuse, and J.~M. Robins (2004).
\newblock Sensitivity analyses for unmeasured confounding assuming a marginal
  structural model for repeated measures.
\newblock {\em Statistics in medicine\/}~{\em 23\/}(5), 749--767.

\bibitem[\protect\citeauthoryear{Chen, Leung, and Qin}{Chen
  et~al.}{2008}]{chen2008improving}
Chen, S.~X., D.~H. Leung, and J.~Qin (2008).
\newblock Improving semiparametric estimation by using surrogate data.
\newblock {\em Journal of the Royal Statistical Society: Series B\/}~{\em
  70\/}(4), 803--823.

\bibitem[\protect\citeauthoryear{Cole and Hern{\'a}n}{Cole and
  Hern{\'a}n}{2008}]{cole2008constructing}
Cole, S.~R. and M.~A. Hern{\'a}n (2008).
\newblock Constructing inverse probability weights for marginal structural
  models.
\newblock {\em American Journal of Epidemiology\/}~{\em 168\/}(6), 656--664.

\bibitem[\protect\citeauthoryear{Cox}{Cox}{2006}]{cox2006principles}
Cox, D.~R. (2006).
\newblock {\em Principles of Statistical Inference}.
\newblock Cambridge University Press.

\bibitem[\protect\citeauthoryear{Dawid}{Dawid}{1979}]{dawid1979conditional}
Dawid, A.~P. (1979).
\newblock Conditional independence in statistical theory.
\newblock {\em Journal of the Royal Statistical Society, Series B\/}, 1--31.

\bibitem[\protect\citeauthoryear{Deville and S{\"a}rndal}{Deville and
  S{\"a}rndal}{1992}]{deville1992calibration}
Deville, J.-C. and C.-E. S{\"a}rndal (1992).
\newblock Calibration estimators in survey sampling.
\newblock {\em Journal of the American statistical Association\/}~{\em
  87\/}(418), 376--382.

\bibitem[\protect\citeauthoryear{Ding and Li}{Ding and
  Li}{2018}]{ding2018causal}
Ding, P. and F.~Li (2018).
\newblock Causal inference: a missing data perspective.
\newblock {\em Arxiv, preprint arXiv:1712.06170\/}.

\bibitem[\protect\citeauthoryear{Ertefaie and Stephens}{Ertefaie and
  Stephens}{2010}]{ertefaie2010comparing}
Ertefaie, A. and D.~A. Stephens (2010).
\newblock Comparing approaches to causal inference for longitudinal data:
  Inverse probability weighting versus propensity scores.
\newblock {\em The International Journal of Biostatistics\/}~{\em 6\/}(2),
  1--22.

\bibitem[\protect\citeauthoryear{Fuller, Loughin, and Baker}{Fuller
  et~al.}{1994}]{fuller1994regression}
Fuller, W.~A., M.~M. Loughin, and H.~D. Baker (1994).
\newblock Regression weighting for the 1987-88 national food consumption
  survey.
\newblock {\em Survey Methodology\/}~{\em 20}, 75--85.

\bibitem[\protect\citeauthoryear{Hern{\'a}n and Robins}{Hern{\'a}n and
  Robins}{2018}]{hernan2018causal}
Hern{\'a}n, M.~A. and J.~M. Robins (2018).
\newblock {\em Causal Inference}.
\newblock Boca Raton: Chapman \& Hall/CRC, forthcoming.

\bibitem[\protect\citeauthoryear{Hirano and Imbens}{Hirano and
  Imbens}{2001}]{hirano2001estimation}
Hirano, K. and G.~W. Imbens (2001).
\newblock Estimation of causal effects using propensity score weighting: An
  application to data on right heart catheterization.
\newblock {\em Health Services and Outcomes Research Methodology\/}~{\em
  2\/}(3-4), 259--278.

\bibitem[\protect\citeauthoryear{Hogan, Roy, and Korkontzelou}{Hogan
  et~al.}{2004}]{hogan2004handling}
Hogan, J.~W., J.~Roy, and C.~Korkontzelou (2004).
\newblock Handling drop-out in longitudinal studies.
\newblock {\em Statistics in Medicine\/}~{\em 23\/}(9), 1455--1497.

\bibitem[\protect\citeauthoryear{Holland}{Holland}{1986}]{holland1986statistics}
Holland, P.~W. (1986).
\newblock Statistics and causal inference.
\newblock {\em Journal of the American Statistical Association\/}~{\em
  81\/}(396), 945--960.

\bibitem[\protect\citeauthoryear{Hong and Raudenbush}{Hong and
  Raudenbush}{2006}]{hong2006evaluating}
Hong, G. and S.~W. Raudenbush (2006).
\newblock Evaluating kindergarten retention policy: A case study of causal
  inference for multilevel observational data.
\newblock {\em Journal of the American Statistical Association\/}~{\em
  101\/}(475), 901--910.

\bibitem[\protect\citeauthoryear{Horvitz and Thompson}{Horvitz and
  Thompson}{1952}]{horvitz1952generalization}
Horvitz, D.~G. and D.~J. Thompson (1952).
\newblock A generalization of sampling without replacement from a finite
  universe.
\newblock {\em Journal of the American Statistical Association\/}~{\em
  47\/}(260), 663--685.

\bibitem[\protect\citeauthoryear{Hosmer and Lemeshow}{Hosmer and
  Lemeshow}{2000}]{hosmer2000applied}
Hosmer, D.~W. and S.~Lemeshow (2000).
\newblock {\em Applied Logistic Regression}.
\newblock John Wiley \& Sons.

\bibitem[\protect\citeauthoryear{Imai and Ratkovic}{Imai and
  Ratkovic}{2014}]{imai2014covariate}
Imai, K. and M.~Ratkovic (2014).
\newblock Covariate balancing propensity score.
\newblock {\em Journal of the Royal Statistical Society, Series B\/}~{\em
  76\/}(1), 243--263.

\bibitem[\protect\citeauthoryear{Imbens and Rubin}{Imbens and
  Rubin}{2015}]{imbens2015causal}
Imbens, G.~W. and D.~B. Rubin (2015).
\newblock {\em Causal Inference for Statistics, Social, and Biomedical
  Sciences: An Introduction}.
\newblock Cambridge University Press.

\bibitem[\protect\citeauthoryear{Kalton and Flores-Cervantes}{Kalton and
  Flores-Cervantes}{2003}]{kalton2003weighting}
Kalton, G. and I.~Flores-Cervantes (2003).
\newblock Weighting methods.
\newblock {\em Journal of Official Statistics\/}~{\em 19\/}(2), 81.

\bibitem[\protect\citeauthoryear{Kim and Im}{Kim and
  Im}{2014}]{kim2014propensity}
Kim, J.~K. and J.~Im (2014).
\newblock Propensity score adjustment with several follow-ups.
\newblock {\em Biometrika\/}~{\em 101\/}(2), 439--448.

\bibitem[\protect\citeauthoryear{Kim and Shao}{Kim and
  Shao}{2013}]{kim2013statistical}
Kim, J.~K. and J.~Shao (2013).
\newblock {\em Statistical Methods for Handling Incomplete Data}.
\newblock CRC Press.

\bibitem[\protect\citeauthoryear{Kott and Chang}{Kott and
  Chang}{2010}]{kott2010using}
Kott, P.~S. and T.~Chang (2010).
\newblock Using calibration weighting to adjust for nonignorable unit
  nonresponse.
\newblock {\em Journal of the American Statistical Association\/}~{\em
  105\/}(491), 1265--1275.

\bibitem[\protect\citeauthoryear{Li, Zaslavsky, and Landrum}{Li
  et~al.}{2013}]{li2013propensity}
Li, F., A.~M. Zaslavsky, and M.~B. Landrum (2013).
\newblock Propensity score weighting with multilevel data.
\newblock {\em Statistics in Medicine\/}~{\em 32\/}(19), 3373--3387.

\bibitem[\protect\citeauthoryear{Little}{Little}{1986}]{little1986survey}
Little, R.~J. (1986).
\newblock Survey nonresponse adjustments for estimates of means.
\newblock {\em International Statistical Review/Revue Internationale de
  Statistique\/}, 139--157.

\bibitem[\protect\citeauthoryear{Liu, Hudgens, and Becker-Dreps}{Liu
  et~al.}{2016}]{liu2016inverse}
Liu, L., M.~G. Hudgens, and S.~Becker-Dreps (2016).
\newblock On inverse probability-weighted estimators in the presence of
  interference.
\newblock {\em Biometrika\/}~{\em 103\/}(4), 829--842.

\bibitem[\protect\citeauthoryear{Lunceford and Davidian}{Lunceford and
  Davidian}{2004}]{lunceford2004stratification}
Lunceford, J.~K. and M.~Davidian (2004).
\newblock Stratification and weighting via the propensity score in estimation
  of causal treatment effects: a comparative study.
\newblock {\em Statistics in Medicine\/}~{\em 23\/}(19), 2937--2960.

\bibitem[\protect\citeauthoryear{Miao, Tchetgen~Tchetgen, and Geng}{Miao
  et~al.}{2015}]{miao2015identification}
Miao, W., E.~Tchetgen~Tchetgen, and Z.~Geng (2015).
\newblock Identification and doubly robust estimation of data missing not at
  random with a shadow variable.
\newblock {\em arXiv preprint arXiv:1509.02556\/}.

\bibitem[\protect\citeauthoryear{Mitra and Reiter}{Mitra and
  Reiter}{2011}]{mitra2011estimating}
Mitra, R. and J.~P. Reiter (2011).
\newblock Estimating propensity scores with missing covariate data using
  general location mixture models.
\newblock {\em Statistics in Medicine\/}~{\em 30\/}(6), 627--641.

\bibitem[\protect\citeauthoryear{Naimi, Moodie, Auger, and Kaufman}{Naimi
  et~al.}{2014}]{naimi2014constructing}
Naimi, A.~I., E.~E. Moodie, N.~Auger, and J.~S. Kaufman (2014).
\newblock Constructing inverse probability weights for continuous exposures: a
  comparison of methods.
\newblock {\em Epidemiology\/}~{\em 25\/}(2), 292--299.

\bibitem[\protect\citeauthoryear{Neyman}{Neyman}{1990}]{splawa1990application}
Neyman, J. (1923, 1990).
\newblock On the application of probability theory to agricultural experiments.
  {E}ssay on principles. {S}ection 9.
\newblock {\em Statistical Science\/}~{\em 5\/}(4), 465--472.

\bibitem[\protect\citeauthoryear{Neyman and Scott}{Neyman and
  Scott}{1948}]{neyman1948consistent}
Neyman, J. and E.~L. Scott (1948).
\newblock Consistent estimates based on partially consistent observations.
\newblock {\em Econometrica: Journal of the Econometric Society\/}~{\em
  16\/}(1), 1--32.

\bibitem[\protect\citeauthoryear{O'Connor, Sorden, and Apley}{O'Connor
  et~al.}{2005}]{o2005association}
O'Connor, A.~M., S.~D. Sorden, and M.~D. Apley (2005).
\newblock Association between the existence of calves persistently infected
  with bovine viral diarrhea virus and commingling on pen morbidity in feedlot
  cattle.
\newblock {\em American Journal of Veterinary Research\/}~{\em 66\/}(12),
  2130--2134.

\bibitem[\protect\citeauthoryear{Ogburn, Rotnitzky, and Robins}{Ogburn
  et~al.}{2015}]{ogburn2015doubly}
Ogburn, E.~L., A.~Rotnitzky, and J.~M. Robins (2015).
\newblock Doubly robust estimation of the local average treatment effect curve.
\newblock {\em Journal of the Royal Statistical Society: Series B (Statistical
  Methodology)\/}~{\em 77\/}(2), 373--396.

\bibitem[\protect\citeauthoryear{Pearl, Glymour, and Jewell}{Pearl
  et~al.}{2016}]{pearl2016causal}
Pearl, J., M.~Glymour, and N.~P. Jewell (2016).
\newblock {\em Causal Inference in Statistics: A Primer}.
\newblock John Wiley \& Sons.

\bibitem[\protect\citeauthoryear{Ramirez, Wang, Prickett, Pogranichniy, Yoon,
  Main, Johnson, Rademacher, Hoogland, Hoffmann, et~al.}{Ramirez
  et~al.}{2012}]{ramirez2012efficient}
Ramirez, A., C.~Wang, J.~R. Prickett, R.~Pogranichniy, K.-J. Yoon, R.~Main,
  J.~K. Johnson, C.~Rademacher, M.~Hoogland, P.~Hoffmann, et~al. (2012).
\newblock Efficient surveillance of pig populations using oral fluids.
\newblock {\em Preventive Veterinary Medicine\/}~{\em 104\/}(3-4), 292--300.

\bibitem[\protect\citeauthoryear{Robins, Hern{\'a}n, and Brumback}{Robins
  et~al.}{2000}]{robins2000marginal}
Robins, J.~M., M.~{\'A}. Hern{\'a}n, and B.~Brumback (2000).
\newblock Marginal structural models and causal inference in epidemiology.
\newblock {\em Epidemiology\/}~{\em 11\/}(5), 550--560.

\bibitem[\protect\citeauthoryear{Rosenbaum and Rubin}{Rosenbaum and
  Rubin}{1983}]{rosenbaum1983central}
Rosenbaum, P.~R. and D.~B. Rubin (1983).
\newblock The central role of the propensity score in observational studies for
  causal effects.
\newblock {\em Biometrika\/}~{\em 70\/}(1), 41--55.

\bibitem[\protect\citeauthoryear{Rotnitzky and Robins}{Rotnitzky and
  Robins}{1995}]{rotnitzky1995semiparametric}
Rotnitzky, A. and J.~M. Robins (1995).
\newblock Semiparametric regression estimation in the presence of dependent
  censoring.
\newblock {\em Biometrika\/}~{\em 82\/}(4), 805--820.

\bibitem[\protect\citeauthoryear{Rubin}{Rubin}{1974}]{rubin1974estimating}
Rubin, D.~B. (1974).
\newblock Estimating causal effects of treatments in randomized and
  nonrandomized studies.
\newblock {\em Journal of Educational Psychology\/}~{\em 66\/}(5), 688--701.

\bibitem[\protect\citeauthoryear{Simpson}{Simpson}{1951}]{simpson1951interpretation}
Simpson, E.~H. (1951).
\newblock The interpretation of interaction in contingency tables.
\newblock {\em Journal of the Royal Statistical Society, Series B\/}, 238--241.

\bibitem[\protect\citeauthoryear{Sj{\"o}lander, Nyr{\'e}n, Bellocco, and
  Evans}{Sj{\"o}lander et~al.}{2011}]{sjolander2011comparing}
Sj{\"o}lander, A., O.~Nyr{\'e}n, R.~Bellocco, and M.~Evans (2011).
\newblock Comparing different strategies for timing of dialysis initiation
  through inverse probability weighting.
\newblock {\em American Journal of Epidemiology\/}~{\em 174\/}(10), 1204--1210.

\bibitem[\protect\citeauthoryear{Skinner and D'arrigo}{Skinner and
  D'arrigo}{2011}]{skinner2011inverse}
Skinner, C. and J.~D'arrigo (2011).
\newblock Inverse probability weighting for clustered nonresponse.
\newblock {\em Biometrika\/}~{\em 98\/}(4), 953--966.

\bibitem[\protect\citeauthoryear{Sun and Tchetgen~Tchetgen}{Sun and
  Tchetgen~Tchetgen}{2017}]{sun2017inverse}
Sun, B. and E.~J. Tchetgen~Tchetgen (2017).
\newblock On inverse probability weighting for nonmonotone missing at random
  data.
\newblock {\em Journal of the American Statistical Association\/}, 1--11.

\bibitem[\protect\citeauthoryear{Tan}{Tan}{2010}]{tan2010bounded}
Tan, Z. (2010).
\newblock Bounded, efficient and doubly robust estimation with inverse
  weighting.
\newblock {\em Biometrika\/}~{\em 97\/}(3), 661--682.

\bibitem[\protect\citeauthoryear{Tchetgen~Tchetgen and
  VanderWeele}{Tchetgen~Tchetgen and VanderWeele}{2012}]{tchetgen2012causal}
Tchetgen~Tchetgen, E.~J. and T.~J. VanderWeele (2012).
\newblock On causal inference in the presence of interference.
\newblock {\em Statistical Methods in Medical Research\/}~{\em 21\/}(1),
  55--75.

\bibitem[\protect\citeauthoryear{Thoemmes and West}{Thoemmes and
  West}{2011}]{thoemmes2011use}
Thoemmes, F.~J. and S.~G. West (2011).
\newblock The use of propensity scores for nonrandomized designs with clustered
  data.
\newblock {\em Multivariate Behavioral Research\/}~{\em 46\/}(3), 514--543.

\bibitem[\protect\citeauthoryear{Tsiatis}{Tsiatis}{2006}]{tsiatis2006semiparametric}
Tsiatis, A. (2006).
\newblock {\em Semiparametric Theory and Missing Data}.
\newblock Springer.

\bibitem[\protect\citeauthoryear{VanderWeele}{VanderWeele}{2009}]{vanderweele2009marginal}
VanderWeele, T.~J. (2009).
\newblock Marginal structural models for the estimation of direct and indirect
  effects.
\newblock {\em Epidemiology\/}~{\em 20\/}(1), 18--26.

\bibitem[\protect\citeauthoryear{Vansteelandt and Daniel}{Vansteelandt and
  Daniel}{2014}]{vansteelandt2014regression}
Vansteelandt, S. and R.~M. Daniel (2014).
\newblock On regression adjustment for the propensity score.
\newblock {\em Statistics in Medicine\/}~{\em 33\/}(23), 4053--4072.

\bibitem[\protect\citeauthoryear{Wen and Seaman}{Wen and
  Seaman}{2018}]{wen2018semi}
Wen, L. and S.~R. Seaman (2018).
\newblock Semi-parametric methods of handling missing data in mortal cohorts
  under non-ignorable missingness.
\newblock {\em Biometrics\/}.

\bibitem[\protect\citeauthoryear{Yang}{Yang}{2017}]{yang2017propensity}
Yang, S. (2017).
\newblock Propensity score weighting for causal inference with clustered data.
\newblock {\em Arxiv, preprint arXiv:1703.06086\/}.

\bibitem[\protect\citeauthoryear{Yuan and Little}{Yuan and
  Little}{2007}]{yuan2007model}
Yuan, Y. and R.~J. Little (2007).
\newblock Model-based estimates of the finite population mean for two-stage
  cluster samples with unit non-response.
\newblock {\em Journal of the Royal Statistical Society: Series C (Applied
  Statistics)\/}~{\em 56\/}(1), 79--97.

\bibitem[\protect\citeauthoryear{Zhang, Tsiatis, Laber, and Davidian}{Zhang
  et~al.}{2012}]{zhang2012robust}
Zhang, B., A.~A. Tsiatis, E.~B. Laber, and M.~Davidian (2012).
\newblock A robust method for estimating optimal treatment regimes.
\newblock {\em Biometrics\/}~{\em 68\/}(4), 1010--1018.

\bibitem[\protect\citeauthoryear{Zhao, Small, and Bhattacharya}{Zhao
  et~al.}{2017}]{zhao2017sensitivity}
Zhao, Q., D.~S. Small, and B.~B. Bhattacharya (2017).
\newblock Sensitivity analysis for inverse probability weighting estimators via
  the percentile bootstrap.
\newblock {\em arXiv preprint arXiv:1711.11286\/}.

\bibitem[\protect\citeauthoryear{Zubizarreta and Keele}{Zubizarreta and
  Keele}{2017}]{zubizarreta2017optimal}
Zubizarreta, J.~R. and L.~Keele (2017).
\newblock Optimal multilevel matching in clustered observational studies: A
  case study of the effectiveness of private schools under a large-scale
  voucher system.
\newblock {\em Journal of the American Statistical Association\/}~{\em
  112\/}(518), 547--560.

\end{thebibliography}
	
\newpage
\appendix	
\begin{center}
	{\Large Supplementary Materials}
\end{center}
\section[A]{Proof of Theorem 1}\label{appx:proof_probs}	
\begin{proof} For any $\boldsymbol{z}\in\Omega_{\boldsymbol{Z}}$ and its corresponding subvector value $  \boldsymbol{z}_{sub}\in\Omega_{\boldsymbol{Z}_{sub}} $, let $ \boldsymbol{z}^* \in \Omega_{\boldsymbol{Z}} $ and its corresponding subvector value $  \boldsymbol{z}_{sub}^*\in\Omega_{\boldsymbol{Z}_{sub}} $ satisfy (i) $ \boldsymbol{z}_{sub} \neq \boldsymbol{z}_{sub}^* $ and (ii) $ T(\boldsymbol{z})=T(\boldsymbol{z}^*) $. 
	Because $ \boldsymbol{T} $ is sufficient for $ \boldsymbol{\theta} $, by Bayes rule,
	\begin{eqnarray*}\label{eq:thm1_frac}
		&&P\{\boldsymbol{Z}_{sub}=\boldsymbol{z}_{sub}|\boldsymbol{X},\boldsymbol{T}=T(\boldsymbol{z})\} \nonumber \\
		&=&P\{\boldsymbol{Z}_{sub}=\boldsymbol{z}_{sub}|\boldsymbol{X},\boldsymbol{\theta}, \boldsymbol{T}=T(\boldsymbol{z})\} \nonumber\\
		&=&\frac{P\{\boldsymbol{Z}_{sub}=\boldsymbol{z}_{sub}, \boldsymbol{T}=T(z)|\boldsymbol{X},\boldsymbol{\theta}\}}{P\{\boldsymbol{T}=T(\boldsymbol{z})|\boldsymbol{X},\boldsymbol{\theta}\}}.  
	\end{eqnarray*}
	Next we want to show the numerator of \eqref{eq:thm1_frac} is in range (0,1). That is, 
	\begin{align*}
	&P\{\boldsymbol{Z}_{sub}=\boldsymbol{z}_{sub}, \boldsymbol{T}=T(z)|\boldsymbol{X},\boldsymbol{\theta}\} \geq P\{\boldsymbol{Z}=\boldsymbol{z}, \boldsymbol{Z}_{sub}=\boldsymbol{z}_{sub}, \boldsymbol{T}=T(z)|\boldsymbol{X},\boldsymbol{\theta}\} = P\{\boldsymbol{Z}=\boldsymbol{z}|\boldsymbol{X},\boldsymbol{\theta}\}>0, \nonumber\\
	&P\{\boldsymbol{Z}_{sub}=\boldsymbol{z}_{sub}, \boldsymbol{T}=T(\boldsymbol{z})|\boldsymbol{X},\boldsymbol{\theta}\} \leq P\{\boldsymbol{Z}_{sub}=\boldsymbol{z}_{sub}|\boldsymbol{X},\boldsymbol{\theta}\} <1.\nonumber
	\end{align*} 
	Therefore, $0<P\{\boldsymbol{Z}_{sub}=\boldsymbol{z}_{sub}, \boldsymbol{T}=T(\boldsymbol{z})|\boldsymbol{X},\boldsymbol{\theta}\}<1$.	Last we want to show the denominator of \eqref{eq:thm1_frac} is greater than the numerator, which is 
	\begin{align*}\label{eq:thm1_inject1}
	P\{\boldsymbol{T}=T(\boldsymbol{z})|\boldsymbol{X},\boldsymbol{\theta}\}&\geq P\{\boldsymbol{Z}_{sub}=\boldsymbol{z}_{sub}, \boldsymbol{T}=T(\boldsymbol{z})|\boldsymbol{X},\boldsymbol{\theta}\} + P\{\boldsymbol{Z}_{sub}=\boldsymbol{z}_{sub}^*, \boldsymbol{T}=T(\boldsymbol{z}^*)|\boldsymbol{X},\boldsymbol{\theta}\}\nonumber\\
	&> P\{\boldsymbol{Z}_{sub}=\boldsymbol{z}_{sub}, \boldsymbol{T}=T(\boldsymbol{z})|\boldsymbol{X},\boldsymbol{\theta}\}.
	\end{align*}
	So we can show that $0< P\{\boldsymbol{Z}_{sub}=\boldsymbol{z}_{sub}|\boldsymbol{X},\boldsymbol{T}=T(\boldsymbol{z})\}<1 $.
\end{proof}	

\newpage
\section{Proof of Theorem 2}\label{appx:proof_indep}
\begin{proof}
	From Lemma 4.2 of \cite{dawid1979conditional}, we know that if $ \boldsymbol{Z}_1\indept \boldsymbol{Z}_2|\boldsymbol{Z}_3,\boldsymbol{\theta} $ and $ \boldsymbol{T} $ is a function of $ \boldsymbol{Z}_1 $, then $ \boldsymbol{T} \indept \boldsymbol{Z}_2|\boldsymbol{Z}_3,\boldsymbol{\theta} $ and $ \boldsymbol{Z}_1\indept \boldsymbol{Z}_2|\boldsymbol{Z}_3,\boldsymbol{\theta},\boldsymbol{T} $. Moreover, $ \boldsymbol{T} $ is sufficient for $ \boldsymbol{\theta} $, then $ P(\boldsymbol{Z}_1|\boldsymbol{Z}_3,\boldsymbol{T})=P(\boldsymbol{Z}_1|\boldsymbol{Z}_3,\boldsymbol{T},\boldsymbol{\theta}) $.
	
	Let $ f(\boldsymbol{\theta}|\boldsymbol{Z}_3, \boldsymbol{T}) $ be the conditional density function for $ \boldsymbol{\theta} $ given $ \boldsymbol{Z}_3 $ and $ \boldsymbol{T} $. If $ \boldsymbol{Z}_1\indept \boldsymbol{Z}_2|\boldsymbol{Z}_3,\boldsymbol{\theta} $, the joint probability for $ \boldsymbol{Z}_1 $ and $ \boldsymbol{Z}_2 $ conditional on $ \boldsymbol{Z}_3 $ and $ \boldsymbol{T} $ is
	\begin{eqnarray*}
		P(\boldsymbol{Z}_1,\boldsymbol{Z}_2|\boldsymbol{Z}_3,\boldsymbol{T})&=&\int_{\Omega_{\boldsymbol{\theta}}}P(\boldsymbol{Z}_1,\boldsymbol{Z}_2,\boldsymbol{\theta}|\boldsymbol{Z}_3,\boldsymbol{T})f(\boldsymbol{\theta}|\boldsymbol{Z}_3,\boldsymbol{T})d\boldsymbol{\theta} \nonumber\\
		&=&\int_{\Omega_{\boldsymbol{\theta}}}P(\boldsymbol{Z}_1,\boldsymbol{Z}_2|\boldsymbol{Z}_3,\boldsymbol{T},\boldsymbol{\theta})f^2(\boldsymbol{\theta}|\boldsymbol{Z}_3,\boldsymbol{T})d\boldsymbol{\theta} \nonumber\\
		&=&\int_{\Omega_{\boldsymbol{\theta}}}P(\boldsymbol{Z}_1|\boldsymbol{Z}_3,\boldsymbol{T},\boldsymbol{\theta})P(\boldsymbol{Z}_2|\boldsymbol{Z}_3,T,\boldsymbol{\theta})f^2(\boldsymbol{\theta}|\boldsymbol{Z}_3,\boldsymbol{T})d\boldsymbol{\theta} \nonumber\\
		&=&\int_{\Omega_{\boldsymbol{\theta}}}P(\boldsymbol{Z}_1|\boldsymbol{Z}_3,\boldsymbol{T})P(\boldsymbol{Z}_2|\boldsymbol{Z}_3,\boldsymbol{T},\boldsymbol{\theta})f^2(\boldsymbol{\theta}|\boldsymbol{Z}_3,\boldsymbol{T})d\boldsymbol{\theta} \nonumber\\
		&=&P(\boldsymbol{Z}_1|\boldsymbol{Z}_3,\boldsymbol{T})\int_{\Omega_{\boldsymbol{\theta}}}P(\boldsymbol{Z}_2|\boldsymbol{Z}_3,\boldsymbol{T},\boldsymbol{\theta})f^2(\boldsymbol{\theta}|\boldsymbol{Z}_3,\boldsymbol{T})d\boldsymbol{\theta} \nonumber\\		
		&=&P(\boldsymbol{Z}_1|\boldsymbol{Z}_3,\boldsymbol{T})\int_{\Omega_{\boldsymbol{\theta}}}P(\boldsymbol{Z}_2,\boldsymbol{\theta}|\boldsymbol{Z}_3,\boldsymbol{T})f(\boldsymbol{\theta}|\boldsymbol{Z}_3,\boldsymbol{T})d\boldsymbol{\theta} \nonumber\\				
		&=&P(\boldsymbol{Z}_1|\boldsymbol{Z}_3,\boldsymbol{T})P(\boldsymbol{Z}_2|\boldsymbol{Z}_3,\boldsymbol{T}). 
	\end{eqnarray*}
\end{proof}

\newpage
\section{Proof of Theorem 3}\label{appx:robust}
\begin{proof}
	For treatment level $ a $, we want to show the term $ E\Big\{\frac{1}{n}\sum_{i=1}^{m}\sum_{j=1}^{n_i}\frac{\mathcal{I}(A_{ij}=a)Y_{ij}}{P(A_{ij}=a|\boldsymbol{X}_{i},T_i;\boldsymbol{\beta})} \Big\} $ is unbiased to $ EY(a) $ and robust to both the correlation between $ U_i $ and $ \boldsymbol{X}_i $ and the correlation between $ U_i $ and $ \boldsymbol{Y}_i(a) $.
	
	We treat $ U_i $ as the cluster-specific parameter for all $ i $. From Lemma 4.2 of \cite{dawid1979conditional}, $ T_i $, a function of $ \boldsymbol{A}_i $, satisfies Assumption 
	5. We have $ \boldsymbol{Y}_i(a) \indept \boldsymbol{A}_i|\boldsymbol{X}_i,T_i, U_i$ for all $ i $ and $ a $. Therefore,
	\begin{align*}
	E_{A_{ij}|\boldsymbol{X}_{i},T_i,U_i}\Big\{\frac{\mathcal{I}(A_{ij}=a)}{P(A_{ij}=a|\boldsymbol{X}_{i}, T_i)}\bigg|\boldsymbol{X}_i, T_i,U_i \Big\} = E_{A_{ij}|\boldsymbol{X}_{i},T_i,U_i}\Big\{\frac{\mathcal{I}(A_{ij}=a)}{P(A_{ij}=a|\boldsymbol{X}_{i}, T_i,U_i)}\bigg|\boldsymbol{X}_i, T_i,U_i \Big\} =1.
	\end{align*}

	\begin{align*}
	&E\Big\{\frac{1}{n}\sum_{i=1}^{m}\sum_{j=1}^{n_i}\frac{\mathcal{I}(A_{ij}=a)Y_{ij}}{P(A_{ij}=1|\boldsymbol{X}_{i},T_i;\boldsymbol{\beta})} \Big\} \nonumber\\
	=&\frac{1}{n}\sum_{i=1}^{m}\sum_{j=1}^{n_i}E_{A_{ij}, Y_{ij}(a)}\Big\{\frac{\mathcal{I}(A_{ij}=a)}{P(A_{ij}=a|\boldsymbol{X}_{i},T_i)}Y_{ij}(a)\Big\}\nonumber\\
	=&\frac{1}{n}\sum_{i=1}^{m}\sum_{j=1}^{n_i}E_{\boldsymbol{X}_{i},T_i,U_i}\left[E_{A_{ij}, Y_{ij}(a)|\boldsymbol{X}_{i},T_i,U_i}\Big\{\frac{\mathcal{I}(A_{ij}=a)}{P(A_{ij}=a|\boldsymbol{X}_{i},T_i,U_i)}Y_{ij}(a)\bigg|\boldsymbol{X}_{i},T_i,U_i \Big\}\right]\nonumber\\
	=&\frac{1}{n}\sum_{i=1}^{m}\sum_{j=1}^{n_i}E_{\boldsymbol{X}_{i},T_i,U_i}\left[E_{A_{ij}|\boldsymbol{X}_{i},T_i,U_i}\Big\{\frac{\mathcal{I}(A_{ij}=a)}{P(A_{ij}=a|\boldsymbol{X}_{i}, T_i,U_i)}\bigg|\boldsymbol{X}_i, T_i,U_i \Big\} E_{Y_{ij}(a)|\boldsymbol{X}_{i},T_i,U_i}\{Y_{ij}(a)|\boldsymbol{X}_{i},T_i,U_i\} \right]\nonumber\\
	=&\frac{1}{n}\sum_{i=1}^{m}\sum_{j=1}^{n_i}E_{\boldsymbol{X}_{i},T_i,U_i}[ E_{Y_{ij}(a)|\boldsymbol{X}_{i},T_i,U_i}\{Y_{ij}(a)|\boldsymbol{X}_{i},T_i,U_i\} ]\nonumber\\
	=&\frac{1}{n}\sum_{i=1}^{m}\sum_{j=1}^{n_i}E_{Y_{ij}(a)}[Y_{ij}(a)] \nonumber\\
	=&E[Y(a)]. 	
	\end{align*}
	The above equation holds due to Assumptions 
	2 and 4*. Therefore, we can prove Theorem 
	3. 
\end{proof}

\section{Conditions for the Asymptotic Properties of CMLE of $ \boldsymbol{\beta} $}\label{appx:asymp.beta}
In \cite{andersen1970asymptotic}, $ \boldsymbol{\beta} $ is called a structural parameter and $ b $'s are called incidental parameters. In our case, we treat $ \{U_i\}_{i=1}^m $ as the incidental parameters and $ U_i\in\Omega_{U} $, which is compact. Moreover, $ n_i $ is fixed and bounded. The following conditions are adopted from those in \cite{andersen1970asymptotic}.

Condition 1: The log density function $ \log \phi_{ij}(a|\boldsymbol{x}_i,t_i;\boldsymbol{\beta}) $ is a differentiable function of $ \boldsymbol{\beta} $ for all $ i $ and $ j $, and there exists a set $ B $ of values of $ t $ with $ P_{\boldsymbol{\beta}_0,U}(T^{-1}B)>0 $ for all $ U\in\Omega_{U} $ and an open set $ \Omega_{0} $ containing the true value of the parameter $ \boldsymbol{\beta}_0 $ such that for any minimal sufficient statistic $ t\in B $ and $ \boldsymbol{x}\in\Omega_{\boldsymbol{X}} $, where $ \Omega_{\boldsymbol{X}} $ is compact, the functions $ \phi_{ij}(a|\boldsymbol{x},t;\boldsymbol{\beta}) $ and $ \phi_{ij}(a|\boldsymbol{x},t;\boldsymbol{\beta}') $ are not identical for any pair $ \boldsymbol{\beta}\in\Omega_{0} $ and $ \boldsymbol{\beta}'\in\Omega_{0} $. 

Condition 2: The maximum likelihood estimating equation
\begin{equation*}\label{appx:eq.cmle}
\sum_{i=1}^{m}\sum_{j=1}^{n_i}\{\partial\log \phi_{ij}(a_{ij}|\boldsymbol{x}_i,t_i;\boldsymbol{\beta})/\partial\boldsymbol{\beta} \} = 0
\end{equation*}
has a unique solution $ \hat{\boldsymbol{\beta}}_n\in \Omega_{\boldsymbol{\beta}} $, which is compact,  for almost all values of the vector $ (t_1, \cdots, t_m) $.

Condition 3: For all $ \boldsymbol{\delta}\in \Delta $, where $ \Delta $ is an open set containing $ \boldsymbol{0} $, 
\begin{equation*}\label{appx:eq.sigma}
\sum_{i=1}^{\infty}\sigma^2(\boldsymbol{\delta},U_i)/i^2 < \infty,
\end{equation*} 
where $ \sigma^2(\boldsymbol{\delta},U_i)=var_{\boldsymbol{\beta}_0,U_i}\{\sum_{j=1}^{n_i}\log \phi_{ij}(A_{ij}|\boldsymbol{X}_i,T_i;\boldsymbol{\beta}_0+\boldsymbol{\delta}) - \sum_{j=1}^{n_i}\log \phi_{ij}(A_{ij}|\boldsymbol{X}_i,T_i;\boldsymbol{\beta}_0) \} $ for all $ U_i\in\Omega_{U} $.

Condition 4: 
The set of first, second, and third partial derivatives of cluster-level log joint density function $ \sum_{j=1}^{n_i}\log \phi_{j}(a|\boldsymbol{X},T;\boldsymbol{\beta}) $ exist for all $ \boldsymbol{\beta} $ in an open set $ \Omega_{0} $ enclosing $ \boldsymbol{\beta}_0 $. 
Let 
\begin{equation*}
B_2(\boldsymbol{\beta})=\frac{1}{n}\sum_{i=1}^{m}\sum_{j=1}^{n_i} E\Big\{\frac{\partial}{\partial\boldsymbol{\beta}\partial\boldsymbol{\beta}^T}\log\phi_{ij}(A_{ij}|\boldsymbol{X}_i,T_i;\boldsymbol{\beta})\Big\}
\end{equation*}
and 
\begin{equation*}
B_3(\boldsymbol{\beta})=\frac{1}{n}\sum_{i=1}^{m} E\Big\{\sum_{j=1}^{n_i}\frac{\partial}{\partial\boldsymbol{\beta}}\log\phi_{ij}(A_{ij}|\boldsymbol{X}_i,T_i;\boldsymbol{\beta})\Big\}\Big\{\sum_{j=1}^{n_i}\frac{\partial}{\partial\boldsymbol{\beta}^T}\log\phi_{ij}(A_{ij}|\boldsymbol{X}_i,T_i;\boldsymbol{\beta})\Big\}.
\end{equation*}
For all $ U\in\Omega_{U} $ and all $ \boldsymbol{\beta} \in \Omega_{0} $, we have
\begin{equation*}\label{appx:cond4.1}
E_{\boldsymbol{\beta},U}\{\sum_{j=1}^{n_i} \partial\log \phi_{j}(A_j|\boldsymbol{X},T;\boldsymbol{\beta})/\partial\beta_k\} =0.
\end{equation*}
for $ k=1,\cdots,p $. There further exist positive integrable functions $ c_{kl}(a_1,\cdots,a_{n_i}) $ such that 
\begin{equation*}\label{appx:cond4.3}
| \sum_{j=1}^{n_i} \partial^3\log \phi_{j}(a_j|\boldsymbol{X},T;\boldsymbol{\beta})/(\partial\beta_k\partial\beta_l\partial\beta_q) | \leq c_{kl}(a_1,\cdots,a_{n_i})
\end{equation*}
for $ \boldsymbol{\beta} \in \Omega_{0} $ and $ q=1,\cdots,p $, and such that $ E_{\boldsymbol{\beta}_0,U}c_{kl}(A_1,\cdots,A_{n_i}) $ and $ var_{\boldsymbol{\beta}_0,U}c_{kl}(A_1,\cdots,A_{n_i}) $ are continuous.

Condition 5: For all $ a $, the density function $ f(a|\boldsymbol{\beta}_0,U) $ is a continuous function of all $ U\in\Omega_{U} $ and for $ k,l=1,\cdots,p $, 
$ var_{\boldsymbol{\beta}_0,U} \{\sum_{j=1}^{n_i} \partial^2\log \phi_{j}(A_j|\boldsymbol{X},T;\boldsymbol{\beta}_0)/(\partial\beta_k\partial\beta_l) \}
$ and 
$ E_{\boldsymbol{\beta}_0,U}\{\sum_{j=1}^{n_i}\partial\log\phi_{ij}(A_{ij}|\boldsymbol{X}_i,T_i;\boldsymbol{\beta})/\partial{\beta_k}\}\{\sum_{j=1}^{n_i}\partial\log\phi_{ij}(A_{ij}|\boldsymbol{X}_i,T_i;\boldsymbol{\beta})/{\partial{\beta}^T_l}\} $
are continuous of $ U $. 
In addition, the matrix $ B_2(\boldsymbol{\beta}) $ is non-singular.

\newpage
\section{Proof of Theorem 6}\label{appx:assym.ya}
\begin{proof}
	For notation convenience, let 
	\begin{align*}
	\hat{Y}_{n}(\hat{\boldsymbol{\beta}};a)&:=\hat{Y}_{ICPW}(a)=\frac{1}{n}\sum_{i=1}^{m}\sum_{j=1}^{n_i}\frac{\mathcal{I}\{A_{ij}=a\}Y_{ij}}{P(A_{ij}=a|\boldsymbol{X}_{i},T_i;\hat{\boldsymbol{\beta}})}=\frac{1}{n}\sum_{i=1}^{m}\sum_{j=1}^{n_i}\frac{\mathcal{I}\{A_{ij}=a\}Y_{ij}}{\phi_{ij}(A_{ij}|\boldsymbol{X}_i,T_i;\boldsymbol{\beta})},\\
	\hat{Y}_{n}(\boldsymbol{\beta}_0;a)&:=\frac{1}{n}\sum_{i=1}^{m}\sum_{j=1}^{n_i}\frac{\mathcal{I}\{A_{ij}=a\}Y_{ij}}{P(A_{ij}=a|\boldsymbol{X}_{i},T_i;\boldsymbol{\beta}_0)}=\frac{1}{n}\sum_{i=1}^{m}\sum_{j=1}^{n_i}\frac{\mathcal{I}\{A_{ij}=a\}Y_{ij}}{\phi_{ij}(A_{ij}|\boldsymbol{X}_i,T_i;\boldsymbol{\beta}_0)}.
	\end{align*}
	We have 
	\begin{align*}
	\sqrt{n}[\hat{Y}_{ICPW}(a) - E\{Y(a)\}] &= \sqrt{n}[\hat{Y}_{n}(\hat{\boldsymbol{\beta}};a) - E\{Y(a)\}] \\
	&=\sqrt{n}[\hat{Y}_{n}(\hat{\boldsymbol{\beta}};a) - \hat{Y}_{n}(\boldsymbol{\beta}_0;a)] + \sqrt{n}[ \hat{Y}_{n}(\boldsymbol{\beta}_0;a) - E\{Y(a)\}].
	\end{align*}
	By Chebyshev's inequality, we can show that	
	\begin{align*}
	P\{ \sqrt{n} | \hat{Y}_{n}(\boldsymbol{\beta}_0;a) - E\{Y(a)\} |>n \}<\frac{\sigma^2_{1,\boldsymbol{\beta}_0}}{n}.
	\end{align*}
	In other words, $ \sqrt{n}[\hat{Y}_{n}(\boldsymbol{\beta}_0;a) - E\{Y(a)\}] \rightarrow 0 $ in probability when $ n \rightarrow \infty $. By Delta method with Taylor series expansion, 
	\begin{align*}
	\hat{Y}_{n}(\hat{\boldsymbol{\beta}};a) &= \hat{Y}_{n}(\boldsymbol{\beta}_0;a) + \frac{\partial \hat{Y}_{n}({\boldsymbol{\beta}};a)}{\partial \boldsymbol{\beta}}\Big|_{\boldsymbol{\beta}=\boldsymbol{\beta}_0} (\hat{\boldsymbol{\beta}}-\boldsymbol{\beta}_0)+ o_p(1) \\
	&=\hat{Y}_{n}(\boldsymbol{\beta}_0;a) -\frac{1}{n}\sum_{i=1}^{m}\sum_{j=1}^{n_i}\frac{\mathcal{I}\{A_{ij}=a\}Y_{ij}}{\phi_{ij}^2(a|\boldsymbol{X}_i,T_i;\boldsymbol{\beta})}\frac{\partial \phi_{ij}(a|\boldsymbol{X}_i,T_i;\boldsymbol{\beta})}{\partial\boldsymbol{\beta}}\Big|_{\boldsymbol{\beta}=\boldsymbol{\beta}_0} (\hat{\boldsymbol{\beta}}-\boldsymbol{\beta}_0) + o_p(1). 
	\end{align*}
	Therefore, we have
	\begin{align*}
	\sqrt{n}\{\hat{Y}_{ICPW}(a) - \hat{Y}_{n}(\boldsymbol{\beta}_0;a)\} \rightarrow N({0}, V_1(\boldsymbol{\beta}_0)) 
	\end{align*} 
	in distribution as $ n\rightarrow\infty $, where $ V_1(\boldsymbol{\beta}_0) = E[ H_1(\boldsymbol{\beta}_0)^T B_1(\boldsymbol{\beta}_0) H_1(\boldsymbol{\beta}_0) ] $, and  
	\begin{equation*}
	H_1(\boldsymbol{\beta}) = \frac{1}{n}\sum_{i=1}^{m}\sum_{j=1}^{n_i}\frac{\mathcal{I}\{A_{ij}=a\}Y_{ij}}{\phi_{ij}^2(a|\boldsymbol{X}_i,T_i;\boldsymbol{\beta})}\frac{\partial\phi_{ij}(a|\boldsymbol{X}_i,T_i;\boldsymbol{\beta})}{\partial\boldsymbol{\beta}}.
	\end{equation*}
	By Slutsky's theorem, we conclude that 
	\begin{align*}
	\sqrt{n}V_1(\boldsymbol{\beta}_0)^{-1/2}(\hat{Y}_{ICPW}(a) - E\{Y(a)\}) \rightarrow N({0},1). 
	\end{align*} 
\end{proof}	

\newpage
\section{Proof of Theorem 7}\label{appx:assym.tau}
\begin{proof}
	For notation convenience, let 
	\begin{align*}
	\hat{\tau}_{n}(\hat{\boldsymbol{\beta}}):=&\hat{\tau}_{ICPW}\\=&\frac{1}{n}\sum_{i=1}^{m}\sum_{j=1}^{n_i}\Big\{\frac{A_{ij}Y_{ij}}{P(A_{ij}=1|\boldsymbol{X}_{i},T_i;\hat{\boldsymbol{\beta}})} - \frac{(1-A_{ij})Y_{ij}}{1-P(A_{ij}=1|\boldsymbol{X}_{i},T_i;\hat{\boldsymbol{\beta}})} \Big\}\\
	=&\frac{1}{n}\sum_{i=1}^{m}\sum_{j=1}^{n_i}\Big\{\frac{A_{ij}Y_{ij}}{\phi_{ij}(\hat{\boldsymbol{\beta}})} - \frac{(1-A_{ij})Y_{ij}}{1-\phi_{ij}(\hat{\boldsymbol{\beta}})} \Big\},\\
	\hat{\tau}_{n}(\boldsymbol{\beta}_0):=&\frac{1}{n}\sum_{i=1}^{m}\sum_{j=1}^{n_i}\Big\{\frac{A_{ij}Y_{ij}}{P(A_{ij}=1|\boldsymbol{X}_{i},T_i;\boldsymbol{\beta}_0)} - \frac{(1-A_{ij})Y_{ij}}{1-P(A_{ij}=1|\boldsymbol{X}_{i},T_i;\boldsymbol{\beta}_0)} \Big\}\\
	&=\frac{1}{n}\sum_{i=1}^{m}\sum_{j=1}^{n_i}\Big\{\frac{A_{ij}Y_{ij}}{\phi_{ij}(\boldsymbol{\beta}_0)} - \frac{(1-A_{ij})Y_{ij}}{1-\phi_{ij}(\boldsymbol{\beta}_0)} \Big\}.\\
	\end{align*}
	We have 
	\begin{align*}
	\sqrt{n}(\hat{\tau}_{ICPW} - \tau) &= \sqrt{n}[\hat{\tau}_{n}(\hat{\boldsymbol{\beta}}) - \hat{\tau}_{n}(\boldsymbol{\beta}_0)] + \sqrt{n}[\hat{\tau}_{n}(\boldsymbol{\beta}_0) - \tau].
	\end{align*}
	By Chebyshev's inequality, we can show that 	
	\begin{align*}
	P\{ \sqrt{n}|\hat{\tau}_{n}(\boldsymbol{\beta}_0) - \tau |>n  \}<\frac{\sigma^2_{2,\boldsymbol{\beta}_0}}{n}.
	\end{align*}
	In other words, $ \sqrt{n}[\hat{\tau}_{n}(\boldsymbol{\beta}_0) - \tau] \rightarrow 0 $ in probability when $ n \rightarrow \infty $. By Delta method with Taylor series expansion, 
	\begin{align*}
	\hat{\tau}_{n}(\hat{\boldsymbol{\beta}}) &= \hat{\tau}_{n}(\boldsymbol{\beta}_0) + \frac{\partial \hat{\tau}_{n}(\boldsymbol{\beta})}{\partial \boldsymbol{\beta}}\Big|_{\boldsymbol{\beta}=\boldsymbol{\beta}_0} (\hat{\boldsymbol{\beta}}-\boldsymbol{\beta}_0)+ o_p(1) \\
	&=\hat{\tau}_{n}(\boldsymbol{\beta}_0) -\frac{1}{n}\sum_{i=1}^{m}\sum_{j=1}^{n_i}\Big\{\frac{A_{ij}Y_{ij}}{\phi^2_{ij}(\boldsymbol{\beta}_0)} + \frac{(1-A_{ij})Y_{ij}}{[1-\phi_{ij}(\boldsymbol{\beta}_0)]^2} \Big\}\frac{\partial\phi_{ij}(\boldsymbol{\beta})}{\partial\boldsymbol{\beta}}\Big|_{\boldsymbol{\beta}=\boldsymbol{\beta}_0} (\hat{\boldsymbol{\beta}}-\boldsymbol{\beta}_0) + o_p(1). 
	\end{align*}
	Therefore, we have
	\begin{align*}
	\sqrt{n}\{\hat{\tau}_{ICPW} - \hat{\tau}_{n}(\boldsymbol{\beta}_0)\} \rightarrow N({0}, V_2(\boldsymbol{\beta}_0)) 
	\end{align*} 
	in distribution as $ n\rightarrow\infty $, where $ V_2(\boldsymbol{\beta}_0) = E[H_2(\boldsymbol{\beta}_0)^T B_1(\boldsymbol{\beta}_0) H_2(\boldsymbol{\beta}_0)]$, and  
	\begin{equation*}
	H_2(\boldsymbol{\beta}) =  \frac{1}{n}\sum_{i=1}^{m}\sum_{j=1}^{n_i}\Big\{\frac{A_{ij}}{\phi_{ij}^2(\boldsymbol{\beta})}+\frac{1-A_{ij}}{[1-\phi_{ij}(\boldsymbol{\beta})]^2}\Big\}Y_{ij}\frac{\partial\phi_{ij}(\boldsymbol{\beta})}{\partial\boldsymbol{\beta}}.
	\end{equation*}
	By Slutsky's theorem, we conclude that 
	\begin{align*}
	\sqrt{n} V_2(\boldsymbol{\beta}_0)^{-1/2}(\hat{\tau}_{ICPW} - \tau) \rightarrow N({0},1). 
	\end{align*} 
\end{proof}	

\end{document}